\documentclass[11pt]{article}

\usepackage{graphics}
\usepackage{graphicx}
\usepackage{amssymb}
\usepackage{amsmath}

\usepackage{geometry}  
\usepackage{mathptmx}      
\usepackage{natbib}
\usepackage{color}

\usepackage{epstopdf}
\usepackage[T1]{fontenc}	

\newcommand{\s}{\mathrm{sign}}
\newcommand{\diag}{\mathrm{diag}}

\usepackage{appendix}
\usepackage[section]{placeins}
\usepackage[bookmarksopen=true,bookmarksnumbered=true]{hyperref}
\usepackage{fix-unnumbered-sections}
\usepackage[labelfont=bf]{caption}
\usepackage[affil-it]{authblk}

\title{Reconstruction of gene regulatory networks from steady state data}
\author[,1]{Arne B.~Gjuvsland\footnote{Email:\href{mailto:arne.gjuvsland@nmbu.no}{arne.gjuvsland@nmbu.no}}}
\author[2]{Erik Plahte}
\affil[1]{Centre for Integrative Genetics, Dept. of Animal and Aquacultural Sciences,\ Norwegian University of Life Sciences, P.O. Box 5003, N-1432 Ås, Norway}
\affil[2]{Centre for Integrative Genetics, Dept. of Mathematical Sciences and Technology, Norwegian University of Life Sciences, P.O. Box 5003, N-1432 Ås, Norway}
\date{}
\begin{document}
\begin{titlepage}

\maketitle
\section*{Abstract} 

Genes are connected in regulatory networks, often modelled by ordinary differential equations. Changes in expression of a gene propagate to other genes along paths in the network. At a stable state, the system's Jacobian matrix confers information about network connectivity. To disclose the functional properties of genes, knowledge of network connections is essential. We present a new method to reconstruct the Jacobian matrix of models for gene regulatory systems from equilibrium protein concentrations. In a recent paper we defined propagation and feedback functions describing how genetic variation at one gene propagates to the other genes in the network and possibly also back to itself. Here we show how propagation and feedback functions provide relations between equilibrium protein levels which are in principle observable, and Jacobi elements which are not directly observable. We establish exact formulae expressing the Jacobian in terms of derivatives of propagation and feedback functions. Approximating these derivatives from perturbed and unperturbed protein levels, we derive formulae for estimating the Jacobian. We apply the method to models of the \emph{Drosophila} segment polarity network and randomly generated gene networks. Genes could be perturbed in two ways: by modifying mRNA degradation rates, or by allele knockout in diploid models. Comparison with the true Jacobians shows that for noiseless data we obtain hit rates close to 100\% in the former case and in the range 80-90\% in the latter. Our method adds to the network interference toolbox and provides a sign estimate of the Jacobian from steady state data, and a value estimate of the Jacobian if protein degradation rates are known. Also the approach identifies some predicted connections as much more reliable than others, and could point to further experiments for resolving uncertainties in the less dependable Jacobian elements.

\textbf{Keywords:} Gene regulatory networks; Reverse engineering; Jacobian; Feedback

\end{titlepage}


\section*{Background} 

Living organisms contain large numbers of complex networks in which  genes, mRNA molecules, protein, metabolites etc. interact to maintain essential functions and to react to a wide range of external impacts and conditions. Genes interact when the protein output of a gene enters into the cell's biochemical system, and through interactions with other chemical species, frequently by long and intricate pathways, influence the expression of other genes by enhancing or inhibiting transcription or modifying translation. Thus a gene is not an independent  object whose functions are only dependent on its internal structure and properties and the external conditions, but is part of a network, reacts to input from other genes in the network and in turn affect other components of the network \citep{Emmert-Streib2011}. Feedback loops and feedforward motifs are important building blocks in gene networks \citep{Alon2007}. 

Systems of ordinary differential equations are frequently used to model the dynamics of  such  networks. In a dynamic system with a stable point, all solutions within its basin of attraction usually decay exponentially towards the stable point, which is then called hyperbolic. Close to it, dynamic trajectories can be computed approximately once the Jacobian matrix $J$ is known.  The Jacobian matrix also confers information about the network structure of the system: all the system's actions and interactions that are operative in the neighbourhood of the stable point, can be read out from its  elements.

For a system for which no validated model exists, a basic question is whether it is possible to obtain information on the  network topology and connections, which are not directly observable, from the concentrations of mRNA and proteins, which are directly observable. In the literature one finds a large number of papers describing different \emph{reverse engineering} methods, see e.g. reviews by \citet{Brazhnik2005,Camacho2007,Cho2007,EmmertStreib2012,Goutsias2007,Ross2008,Stark2003a,Stark2003b,Tirosh2011,Yip2010,Chai2014}.  There are essentially two main classes of methods: using time series data, or  equilibrium concentrations. The present paper belongs to the latter class. 

By repeatedly perturbing the system to induce a shift of equilibrium values of the state variables, one may hopefully be able to infer the Jacobian matrix elements. The observation that expression of gene $B$ is affected by a perturbation of gene $A$ does not in itself say much about $J$. From this fact alone one cannot tell whether the effect is direct or mediated by one or several other genes. This enigma might be solved by performing more perturbations, but by  an \emph{ad hoc} procedure one soon gets lost. A systematic approach is necessary. 

The present paper is a continuation of our recent analysis of propagation of genetic variation in diploid networks \citep{Plahte2013}. There we developed a formalism for describing how a change of genotype of one gene in a gene regulatory network propagates to other, downstream genes and possibly also back to the gene itself, and how this propagation effect is related to the  structure of the network. 

In the present paper we consider the opposite situation in which so far no network model for some gene regulatory system exists. How, and to what extent, can allele knockouts or other  perturbations  yield information by which a model of the network can be constructed? 

In the following the term ``gene'' should be considered as a \emph{functional module}, ``an entity of known/unknown genes, proteins or metabolites, grouped together and internally connected by complex physico-chemical interactions'' \citep{Yalamanchili2006}. A module may take inputs from many other modules, including itself, but we assume each module is delimited such that it only produces a single output. The process of obtaining information about the local interactions, i.e.~the direct effect of a perturbation of one module on another, from the global effects that result from the web of network connections between the modules, has been called \emph{Modular Response Analysis (MRA)} \citep{Sontag2008}, and was developed by \citet{Kholodenko2002} (see also \citet{Andrec2005,Cho2005,Yalamanchili2006}). Albeit similar, our approach differs from the orginal MRA in several respects. In the original MRA, the authors were only able to determine the rows of the Jacobian up to a scalar multiple. This shortcoming is related to the fact that the equilibrium conditions are unchanged if each rate function is multiplied with a nonzero constant, while the Jacobian is not. However, if the MRA is supplemented by non-steady state data, the full Jacobian can be estimated \citep{Sontag2004}. Using only steady state data, we are able to determine the correct sign of the elements of the Jacobian, and if the protein degradation rates are known, their numeric values as well. 

We consider two particular approaches to modify or perturb a gene and by that modifying the functioning of the network. The first approach involves perturbing the mRNA degradation rates, for instance through RNA interference methods. The second approach is to knock out one of the two alleles of a diploid gene. We illustrate both approaches by \emph{in silico} experiments. We show that if the protein degradation rates can be measured or estimated, both approaches can be used to infer the Jacobian from unperturbed and perturbed protein concentration data from which the Jacobian of the system can be inferred.

\section*{Mathematical analyses and results}
\label{results}

\subsection*{Model framework of gene regulatory networks}
\label{modelframework}
\label{propagationfunctions}

We consider a set of genes believed to be part of a network $\mathcal N$ of $n$ nodes or loci $X_{i}$, $i \in N$, where $N = \{1, \ldots, n\}$ and $n \geq 2$. A real gene regulatory network can be modelled by a dynamic model designed according to the following lines. A non-negative variable $z_{i}$  describes the concentration or amount of the output of $X_{i}$, its time rate of change being given by
\begin{equation}
   \dot{z}_{i} = r_{i}(z,a_{i})  - \gamma_{i}z_{i}, \quad i \in N,
\label{e:model}
\end{equation}
where $z = [z_{1}, \ldots, z_{n}]$. The  differentiable rate function $r_{i}(z,a_{i})$  represents the production rate or dose-response function of $X_{i}$, and $\gamma_{i}$ is its constant relative degradation rate. The quantity $a = \{a_{i}\}$, $i \in N$, represents a set of parameters defining the system's genotype, the subset $a_{i}$ defining the genotype of $X_{i}$. This model framework is commonly used to model gene regulatory networks. In fact, it has been around for decades \citep{deJong2002}. Frequently, the dose-response functions are modelled by means of sigmoidal functions, for example the well-known Hill function. We assume that Eqs.~\eqref{e:model} have a single, hyperbolic equilibrium point $x$. Our goal is to estimate the Jacobian $J$ of Eqs.~\eqref{e:model} in $x$ in terms of experimentally observable quantities.

\textbf{Notation:} Vector and matrix components are indicated by subscripts as usual. Superscripts are used extensively as indices and except in a few obvious cases, never indicate a power. For vectors and matrices, superscripts in parentheses indicate that the enclosed component has been excluded. A superscript $(kj)$ to a matrix, for example  $A^{(kj)}$, indicates that row number $k$ and column number $j$ in $A$ have been deleted. Similarly, $a^{(k)}$ is the set (or vector) $a$ with $a_{k}$ deleted, $a^{(k)} = a \setminus \{a_{k}\}$.  Superscripts in brackets indicate the value when a node has been perturbed. For example, $x_{j}^{[k]}$ is the equilibrium value of $X_{j}$ when node $X_{k}$ has been perturbed. We also use a set notation for subscripts. For example, if $L$ is a subset of $N$, then $x_{L}$ is the vector with components $x_{l}$, $l \in L$. If $A$ is a matrix, $A_{i:}$ denotes row number $i$ in $A$.  The equilibrium condition of $X_{j}$ is denoted by $\mathrm E_{j}$. The Jacobian of Eq.~\eqref{e:model} is $J$, $D = \det(J)$ and $D^{(ij)} = \det(J^{(ij)})$. If $M$ is a square matrix, $\diag(M)$ is the diagonal matrix with the same main diagonal as $M$. The superscript $\mathsf T$ to a matrix denotes its transpose. 

Eq.~\eqref{e:model} could be seen as a simplification of a more realistic regulatory system model comprising mRNA, proteins and metabolites.Often gene outputs do not act directly as transcription factors regulating the expression rates of the genes. Rather, there are frequently long and complicated pathways that propagate and modify the regulatory processes. A philosophy behind Eq.~\eqref{e:model} is that all these complicated reactions can be condensed into the response functions $r_{i}$ \citep{Brazhnik2002}. 
The segment polarity network model of \citet{vonDassow2000} is an example of a model framework in which the concentrations of mRNA and protein for each gene in the network are modelled independently. This model framework, which has been used by a number of authors (see e.g.~\citet{Lewis2003,Ichinose2008,Polynikis2009}), is 
\begin{equation}
    \begin{split}
         \dot{P}_{i} & = \rho_{i} m_{i} - \gamma_{i} P_{i},\\
         \dot{m}_{i} & =  R_{i}(P) - \mu_{i} m_{i},
            \end{split}
    \label{e:fullsystem}
\end{equation}
$i \in N$. Here $P_{i}$ and $m_{i}$ are the concentrations of protein and mRNA of gene number $i$, respectively, $R_{i}$ is the production rate (dose-response function) of mRNA, dependent on the concentration of the input proteins,  $\rho_{i}$ is the mRNA-protein conversion rate, and $\gamma_{i}$ and $\mu_{i}$ are positive relative degradation rates. The gene products might act directly as transcription factors, or the function $R_{i}(P)$ might implicitly contain chains of reactions from the gene products to the real transcription factors so that $R_{i}$ is the combined effect of these chains and the transcription. 

Of course the mRNA-protein conversion rate might not be constant, but some nonlinear function of $m$. However, our analysis is local around the stable point $x$, and the second of Eqs.~\eqref{e:fullsystem} would then represent a locally valid linear approximation to the nonlinear transcription model. 

As  mRNA molecules are in general less stable than the corresponding protein molecules, we can safely assume that for all $i$, $\gamma_{i} \ll \mu_{i}$. With a number of reasonable assumptions we can make the quasi-stationarity hypothesis $\dot m_{i} \approx 0$, $m_{i} = R_{i}(x)/ \mu_{i}$. 
This can be justified in a rigorous way by means of singular perturbation theory (see Appendix \ref{mrnaprotein}), and leads to \emph{the reduced model} 
\begin{equation}
\dot{z}_{i} = \frac{\rho_{i}} {\mu_{i}}R_{i}(z) - \gamma_{i} z_{i},
\label{e:reduced}
\end{equation}
where $z = P$. This equation is of the general form Eq.~\eqref{e:model} on which all our derivations are based. For our purposes, Eq.~\eqref{e:reduced} is equivalent to Eq.~\eqref{e:fullsystem}.

In a diploid organism the transcriptional machinery of each gene $X_{i}$ is composed of two alleles, each allele residing in one of the two chromosomes and transcribing mRNA at a certain rate which depends on the genotype and the concentrations of the gene's active transcription factors. Due to small differences in the two alleles' nucleotide sequences, transcription in the two alleles may proceed at (slightly) different transcription rates. The regulatory properties of the two products might also be different. However, a number of experimental results indicate that in many cases, the two alleles of a gene differ only in their regulatory domain without any variation in the coding region \citep{Capon2004,Chamary2005,Duan2003,Gehring2001,Hoogendoorn2003,Jones2012,Mayo2006,Peng2005,Wang1999,Rosenfeld2005}.  In particular it seems reasonable to assume that for a homozygous gene, the two identical alleles are regulated in the same way and produce identical mRNAs.
 
Let the amounts or concentrations of mRNA produced by the two chromosomes of gene $X_{i}$ be $m_{i}^{1}$ and $m_{i}^{2}$, respectively. The total amount (concentration) of mRNA is $m_{i} = m_{i}^{1} + m_{i}^{2}$. The same modelling approach as the one leading to Eq.~\eqref{e:fullsystem} then gives
\begin{equation}
\begin{split}
     	\dot P_{i} & = \rho_{i}m_{i} - \gamma_{i}P_{i},\\
	\dot m_{i}^{1} & = R_{i}(P) - \mu_{i}m_{i}^{1},\\
	\dot m_{i}^{2} & = R_{i}(P) - \mu_{i}m_{i}^{2},
\end{split}
\label{e:pmrna}
\end{equation}
where $P$ is the vector of protein concentrations $P_{i}$. By addition this leads to
\begin{equation}
\begin{split}
     	\dot P_{i} & = \rho_{i}m_{i} - \gamma_{i}P_{i},\\
	\dot m_{i} & = 2R_{i}(P)  - \mu_{i}m_{i}.
\end{split}
\label{e:agg}
\end{equation}

Applying the quasi-stationarity hypothesis to mRNA production leads finally to a single equation for the protein output concentration of node $X_{i}$:
\begin{equation}
   \dot z_{i}  = 2\frac{\rho_{i}}{\mu_{i}}R_{i}(z) - \gamma_{i}z_{i},
\label{e:reduced2}
\end{equation}
where again $z = P$. Eq.~\eqref{e:reduced2}, or alternatively Eq.~\eqref{e:reduced}, is our final model for which we want to estimate the Jacobian $J$.  The stationarity condition of Eq.~\eqref{e:reduced2},
\begin{equation}
  2\frac{\rho_{i}}{\mu_{i}}R_{i}(x) - \gamma_{i}x_{i} = 0,
\label{e:statxi}
\end{equation}
will be denoted $\mathrm E_{i}$ as in \citep{Plahte2013}.

\subsection*{Propagation functions, feedback functions and the Jacobian}
\label{propfeedjac}  

For an investigation of how genetic variation at a locus propagates to the other loci in the network, it is easier and more fruitful to express all equilibrium values $x_{j}$ as functions of the equilibrium value $x_{k}$ of the perturbed node than to express them by the values of perhaps unknown parameters. We showed in \citet{Plahte2013} that the stationarity conditions of Eq.~\eqref{e:model} define $n(n-1)$ \emph{propagation functions} $p_{jk}$ expressing how a change of the equilibrium value $x_{k}$ for the locus $X_{k}$ propagates via the network connections to any other locus $X_{j}$. The relation
 \begin{equation}
   x_{j} = p_{j k}(x_{k},a^{(k)}), \quad j \neq k,
\label{e:psilk}
\end{equation}
where $a^{(k)}$ is the set of parameters not specific to $X_{k}$, expresses an important property of $p_{jk}$; for a given $k$, the propagation functions $p_{jk}$ are invariant under genetic variation of $X_{k}$ \citep{Plahte2013}. 

To determine the changes in any $x_{j}$, $j \neq k$, due to modification imposed on $X_{k}$, all we need is the resulting change in $x_{k}$ and the propagation function $p_{jk}$. We do not need a model for how a genotypic change in $X_{k}$ affects the rate function $r_{k}(z,a_{k})$. This important property is a consequence of a theoretical result \citep{Radulescu2006} stating that the propagation function $p_{j k}(x_{k},a^{(k)})$ is defined by all $\mathrm E_{i}$ except $\mathrm E_{k}$, which is the only one containing the  parameters $a_{k}$ specific to $X_k$.  It follows that the derivative $q_{jk}$ of $p_{jk}$ with respect to $x_{k}$ can be easily estimated by the ratio of a small change in $x_{j}$ divided by the change in $x_{k}$ due to a small genotypic variation in $X_{k}$. On the other hand, we showed in \citet{Plahte2013} that $q_{jk}$ can be expressed in terms of the elements of the Jacobian $J$: 
\begin{equation}
   q_{j k}(x_{k},a^{(k)}) = (-1)^{j + k}\ \frac{D^{(k j)}}{D^{(kk)}}, \quad j \neq k.
\label{e:qjk}
\end{equation}
Accordingly, the propagation functions provide links between the observable protein concentrations and the Jacobian.

Because there are only $n(n-1)$ propagation functions, Eq.~\eqref{e:qjk} alone is not sufficient to determine the $n^{2}$ elements of $J$ completely from observable quantities. It does not tell how a genetic variation of $X_{k}$ affects $x_{k}$ itself. To determine this is a more difficult task because it requires knowledge of how a change of a parameter in $a_{k}$ directly influences $x_{k}$, which in general may require a detailed model of the transcription and translation process. This information is in principle contained in the so-called \emph{feedback functions} defined in \citet{Plahte2013}.

Let $K = N \setminus \{k\}$. Expressing all $x_{j}$, $j \in K$, as $x_{j} = p_{j k}(x_{k},a^{(k)})$ by means of all $\mathrm E_{K}$ and inserting this into $\mathrm E_{k}$ gives 
\begin{equation}
  \gamma_{k} x_{k} =  r_{k}(x_{k},\{p_{jk}(x_{k},a^{(k)})\}_{j \neq k},a_{k}).
\label{e:xk1}
\end{equation}
The right-hand side of this equation is what we call the feedback function $\phi_{k}$ for $X_{k}$. The stationarity condition for $X_{k}$ is then
\begin{equation}
    \gamma_{k}x_{k} =  \phi_{k}(x_{k},a).
\label{e:xk2}
\end{equation}
The feedback function $\phi_{k}$ describes and quantifies the feedback effects of changes in the equilibrium value of $X_{k}$ on itself. If $\psi_{k}(x_{k},a) = \phi_{k}'(x_{k},a) \neq 0$, where the prime denotes the derivative with respect to $x_{k}$, there is an effective feedback of $X_{k}$ on itself, mediated by one or more feedback loops. We showed in \citet{Plahte2013} that 
\begin{equation}
   \psi_{k}(x_{k},a)   =  \frac{D}{D^{(kk)}} + \gamma_{k}.
\label{e:omegak}
\end{equation} 
Combining Eqs.~\eqref{e:qjk} and \eqref{e:omegak} with the well-known formula for the matrix inverse in terms  of determinant and minors, we get for $j \neq k$
\begin{equation}
   (J^{-1})_{jk} = (-1)^{j + k} \frac{D^{(kj)}}{D} = q_{jk}\frac{1}{\psi_{k} - \gamma_{k}},
\label{e:jinvlk}
\end{equation}
or 
\begin{equation}
   JQ = C,
\label{e:jgq}
\end{equation}
where $Q$ is a square matrix defined by $Q_{ij} = q_{ij}$, and $C$ is the diagonal matrix with diagonal elements $\psi_{k} - \gamma_{k}$, $k \in N$. It is easy to see that Eq.~\eqref{e:jinvlk} is valid for $j = k$ as well because $q_{kk} = 1$. 

Let us for a moment assume that $Q$ is known and invertible while $J$ and $C$ are unknown. The effect of $C$ is to multiply each row in $Q^{-1}$ by a constant. Let $c = [c_{1},\ldots, c_{n}]$ be the nonzero diagonal elements of $C$. Because the system  $1$ given by $\dot z_{i} = c_{i}f_{i}(z)$, $i \in N$, has the same stationary states as system $2$ given by $\dot z_{i} = f_{i}(z)$, while their Jacobians are related by $J_{1} = CJ_{2}$, it might seem impossible to determine $C$ from equilibrium values alone (see e.g.~\citet{Sontag2008}). 

The problem is related to the fact that the invariance property of $p_{jk}$ with respect to genotypic variation in $X_{k}$ is not shared by the feedback function $\phi_{k}$. The function itself changes, and unless this dependence is known, Eq.~\eqref{e:xk2} cannot be used to estimate $\psi_{k}(x_{k})$. Accordingly, it is not trivial to obtain an estimate of $\psi_{k}$ by an arbitrary genotypic variation of $X_{k}$. If one were able to change some node-specific parameter in $X_{k}$, the effect on $x_{k}$ would  depend explicitly on this parameter in a way that would require a model for the transcription and translation of the gene.

While our main object is to develop methods to estimate $J$, we note in passing that if $J$ were known, Eq.~\eqref{e:jinvlk} would yield an expression for $Q$. Because all $Q_{kk} = 1$, it follows that $C = (\diag(J^{-1}))^{-1}$, hence
\begin{equation}
   Q = J^{-1}(\diag(J^{-1}))^{-1}.
\label{e:q}
\end{equation}
This formula expresses the total derivative of any variable with respect to any other in terms of the partial derivatives of the dose-response functions and the degradation rates, taking all the network connections into account. This can be interpreted as follows: For the set of differentiable functions $f_{i}: R^{n} \to R^{n}$, $i = 1,\ldots,n$, assume the set of equations $f_{i}(z) = 0$ has a solution $x$,  and let its Jacobian be $J$, $J_{ij} = \partial f_{i}/\partial z_{j}|_{z = x}$.  For each $k$, assume that the set of all the equations except $f_{k}(z) = 0$ define all $x_{j}$ except $x_{k}$ in terms of $x_{k}$ in an open domain around $x$. (If all $f_{j}$ have the particular form assumed in the present paper, this is ensured if a few additional conditions are fulfilled \citep{Radulescu2006}.) Then for any $j,k$, $\mathrm dx_{j}/\mathrm dx_{k} = Q_{jk}$ is given by Eq.~\eqref{e:q}. The formula could be useful for computing the derivates of functions defined implicitly by a set of equations.

 Eq.~\eqref{e:jgq} is the basis for reconstructing the Jacobian from observable equilibrium data. The main problem is to find a way to estimate $C$. Below we present  two different approaches. One is to perturb the degradation rate $\gamma_{k}$ which does not enter into $\phi_{k}$, but enters into Eq.~\eqref{e:xk2} in a known and simple way. The other approach is by allele knockout in diploid, homozygous genes, in which case we may assume that knocking out one of the alleles reduces the production rate of the gene by 50\%.

\subsection*{Reconstruction of $J$  by perturbing the mRNA degradation rates}
\label{gammaperturb}

We first analyse the problem of estimating $J$ by perturbing the mRNA degradation rates $\mu_{i}$ in Eqs.~\eqref{e:reduced} and  \eqref{e:reduced2} and recording the effects on the equilibrium concentrations. Note that $\mu_{i}$ occurs in the dose-response function of $x_{i}$ in a known way. This will lead to an estimate $\widehat J$ of $J$. 

Selecting one node $X_{k}$ and keeping all other parameters fixed, we perturb the degradation rate $\mu_{k}$ from $\mu_{k}$ to $(1 + \omega)\mu_{k}$ and record the unperturbed equilibrium values $x_{j}$ and the perturbed values $x_{j}^{[k]}$. The index in the bracket indicates which node has been perturbed.   Because $\mu_{k}$ occurs in $\phi_{k}$ in a known way, we are able to derive a formula for $\psi_{k}$ for any given $k$.
Again we let $K = N \setminus \{k\}$.
Expressing $x_{K}$ in terms of their propagation functions $p_{Kk}$ (which are defined by all $\mathrm E_{j}$ except $\mathrm E_{k}$), the equation
\begin{equation}
   \gamma_{k}x_{k} = \frac{\rho_{k}} {\mu_{k}}R_{k}(x_{k},p_{Kk}(x_{k})) = \phi_{k}(x_{k})
\label{e:ekmuk}
\end{equation}
defines $x_{k}$ implicitly as a function of $\mu_{k}$.  Implicit differerentiation with respect to $\mu_{k}$ gives readily
\begin{equation*}
   \frac{1} {\psi_{k} - \gamma_{k}} = \frac{\mu_{k}} {\gamma_{k}x_{k}}\frac{\mathrm dx_{k}} {\mathrm d\mu_{k}}.
\end{equation*}
Combined with Eq.~\eqref{e:jinvlk} this yields
\begin{equation}
   (J^{-1})_{jk} = \frac{\mu_{k}} {\gamma_{k}x_{k}}\frac{\mathrm d x_{k}} {\mathrm d\mu_{k}}q_{jk}.\label{e:dxkdmuk}
\end{equation}
Changing $\mu_{k}$ to $(1 + \omega)\mu_{k}$ induces a change from $x_{k}$ to $x_{k}^{[k]}$.
If $|\omega|$ and its effect on $x_{k}$ are small, we can approximate the derivatives by 
\begin{equation}
   \frac{\mathrm d x_{k}} {\mathrm d\mu_{k}} \approx -\frac{\delta_{k}^{[k]}} {\omega\mu_{k}}, \quad q_{jk} \approx \frac{\delta_{j}^{[k]}}{\delta_{k}^{[k]}}, 
\label{e:psiprimeest}
\end{equation}
where $\delta_{j}^{[k]} = x_{j} - x_{j}^{[k]}$ and $\delta_{k}^{[k]} = x_{k} - x_{k}^{[k]}$ are the effects  on $x_{j}$ and $x_{k}$ of perturbing the degradation rate of $X_{k}$.
This gives
\begin{equation}
   (\widehat J^{-1})_{jk} = - \frac{1} {\omega \gamma_{k}x_{k}}{\delta_{j}^{[k]}},
\label{e:hatjmu}
\end{equation}
\begin{equation}
   \widehat J H= - \omega B,
\label{e:jest1}
\end{equation}
where $\widehat J$ is the estimated Jacobian, $H$ is the square matrix with elements $H_{j k} = \delta_{j}^{[k]}$, and $B$ is the diagonal matrix with diagonal elements $\gamma_{k}x_{k}$.  If the $\mu_{k}$ are perturbed by different values $\omega_{k}$, we  get 
\begin{equation}
   \widehat J H= - \Omega B,
\label{e:jest1b}
\end{equation}
where $\Omega$ is the diagonal matrix with $\Omega_{kk} = \omega_{k}$.

A convenient property of  Eqs.~\eqref{e:jest1} and \eqref{e:jest1b} is that if the sign of each $\omega_{k}$ is known, the sign of the elements in $\widehat J$ can be determined even if the values of the $\gamma_{k}$ are unknown, because all diagonal elements in $B$ are positive and $\s(\Omega)$ would be known. In other words,
\begin{equation}
   \s(\widehat{J}) = -\s(\Omega)\s(H^{-1}).
\label{e:sj}
\end{equation}

The advantage of this approach from the mathematical point of view is that the degradation rates can be perturbed by different and small amounts. If there is no noise in the data, the induced errors when derivatives are approximated by ratios of finite differences can be made arbitrarily small. Nevertheless, errors in $\widehat J$ may occur if some nonzero elements in $J$ are very small, or if $H$ is very close to singular. In the latter case, arbitrarily large errors may occur in $\widehat J$ no matter how small $\omega$ is. 

If recordings for several perturbed values of $\mu_{k}$ can be obtained, more precise estimates of the derivatives  could be computed by more advanced mathematical methods. This is a great advantage  of this method compared to the allele knockout method considered in the next subsection.

\subsection*{Reconstruction of $J$ by allele knockouts in diploid loci}
\label{diploidloci}

The allele knockout method is based on the reasonable assumption that if one of the alleles is knocked out, the production rate of the gene for  fixed amounts of its regulators is reduced to one half its unperturbed value. Our starting point is again Eq.~\eqref{e:jinvlk}, where $J$ now is the Jacobian of the system Eq.~\eqref{e:reduced2}. However, in the derivation of \eqref{e:jinvlk} $\phi_{k}$ is derived from $r_{k}$, the dose-response function for a single allele.  As the dose-response function for the homozygous diploid gene $X_{k}$ is $2r_{k}(z)$, Eq.~\eqref{e:jinvlk} must be replaced by 
\begin{equation}
   (J^{-1})_{jk} = (-1)^{j + k} \frac{D^{(kj)}}{D} = q_{jk}\frac{1}{2\psi_{k} - \gamma_{k}}.
\label{e:jinvlkdipl}
\end{equation}
To find an approximation to $\psi_{k} = \mathrm d \phi_{k}/\mathrm dx_{k}$ we use that the unperturbed and knock-out values $x_{k}$ and $x_{k}^{[k]}$ are the solutions of
\begin{equation}
\begin{split}
     	\gamma_{k}x_{k} & = 2\phi_{k}(x_{k}),\\
	\gamma_{k}x_{k}^{[k]} & = \phi_{k}(x_{k}^{[k]}).
	\end{split}
\label{e:singlekocond}
\end{equation}
Eqs.~\eqref{e:singlekocond} give
\begin{equation*}
   \gamma_{k}\delta_{k}^{[k]} = 2\phi_{k}(x_{k}) - \phi_{k}(x_{k}^{[k]}) = 2\phi_{k}(x_{k}) - \phi_{k}(x_{k} - \delta_{k}^{[k]}).
\end{equation*}
If $\delta_{k}^{[k]}$ is small compared to $x_{k}$, expanding the last term to first order leads to
\begin{equation}
   \frac{1} {2\psi_{k} - \gamma_{k}} = -\frac{1} {\gamma_{k}x_{k}^{[k]}}\delta_{k}^{[k]}.
\label{e:psikgammak}
\end{equation}

Combined with Eq.~\eqref{e:psiprimeest} and Eq.~\eqref{e:jinvlkdipl}  this finally gives 
\begin{equation}
   (\widehat{\mspace{1mu}J}^{-1})_{jk} =  -\frac{1}{\gamma_{k}x_{k}^{[k]}}\delta_{j}^{[k]},
\label{e:estjinvlk}
\end{equation}
\begin{equation}
   \widehat J H= -\widetilde B,
\label{e:jest2}
\end{equation}
where $\widetilde B$ is the diagonal matrix with diagonal elements $\gamma_{k}x_{k}^{[k]}$, and $H$ was defined in the previous subsection. Note the similarity between Eqs.~\eqref{e:jest1} and \eqref{e:jest2}. Also note that even if the protein degradation rates are unknown, the dummy values $\gamma_{k} = 1$ will give the same sign to the elements of $\widehat J$ as Eq.~\eqref{e:jest2}.

\subsection*{Conditions on $J$}
\label{signconditions}

One should check that each eigenvalue of $\widehat J$ has a negative real part. When $J$ is estimated by allele knockout, a further condition on the estimated Jacobian follows from \citet{Plahte2013}. For a homozygous locus $X_{k}$ the allele interaction value \citep{Gjuvsland2010} is defined by
\begin{equation}
   \Delta_{k} = x_{k} - 2x_{k}^{[k]}.
\label{e:aivalue}
\end{equation}
Let $F_{k}$ be the sum of all the terms in $D$ in which there is a real regulation of $X_{k}$. This definition inplies that $F_{k}$ does not contain $\gamma_{k}$, rather, each term in $F_{k}$ contains a factor representing a regulation of $X_{k}$, i.e.~a factor $\partial r_{k}/\partial x_{j}$ for some $j$. Each term in $F_{k}$ is a loop product of a feedback loop (called circuit in \citet{Plahte2013}) in $J$. Then 
\begin{equation}
   (-1)^{n}F_{k}\Delta_{k} < 0.
\label{e:fdelta}
\end{equation}

If $P$ is any of these loop products, its contribution to $F_{k}$ is $(-1)^{\nu}P$, where the signature factor $\nu$ is the number of subloops in $P$ with an even number of elements. If $(-1)^{\nu}P$ for some loop has the same sign as $F_{k}$, the loop is sign dominant, and  
\begin{equation}
   (-1)^{n +\nu - 1}P\Delta_{k} > 0.
\label{e:pdelta}
\end{equation}
An estimate $\widehat F_{k}$ of $F_{k}$ can be computed by analysing the loop structure of $\widehat J$, and $\Delta_{k}$ can be computed directly from the observed equilibrium values. If the signs do not match with Eq.~\eqref{e:fdelta} or Eq.~\eqref{e:pdelta}, there could be an error in the loop structure of $\widehat J$, or  the sign of $F_{k}$ or $\Delta_{k}$ could be wrong due to noise. It is not obvious how to extract the most useful and reliable information from these inequalities. In the present paper we have made no effort to check our simulation results agains these sign rules.

\subsection*{Discrepancy measures of estimated Jacobians}
\label{discrepancymeasure}

In tests of the method on systems with a known true Jacobian $J$, the estimate $\widehat J$ could be compared numerically to $J$ in many ways to produce a measure of how well $J$ has been reconstructed. Due to the degradation terms in the rate functions, the diagonal elements of $J$ are generally nonzero except in the unlikely case that a positive autoregulation cancels the degradation term $-\gamma_{j}$. 
Because our method presupposes that the $\gamma_{j}$ are known, we can in fact decide whether a nonzero diagonal element in $\widehat J$ indicates an autoregulation or only linear degradation. For this reason, it is more informative to compare $\widehat K = \widehat J + \Gamma$ with $K = J + \Gamma$, where $\Gamma$ is the diagonal matrix with $\Gamma_{jj} = \gamma_{j}$. 

The choice of error measure should reflect the main purpose of the reconstruction. In our view, the main objective is to reveal the connectivity of the gene  network, while the dynamic properties are of less interest because the true system is most likely highly nonlinear. Our prime concern is therefore whether $\widehat J$ reproduces the right sign structure of $J$, using the sign function $\s(x) = x/|x|$ if $x\neq 0$, $\s(0) = 0$. Predicting the correct numeric magnitude of the matrix elements comes as a subordinate goal. We used the following classification for an estimated element $\widehat{K}_{ij}$ when the true value $K_{ij}$ is known. If $\widehat{K}_{ij}=0$, it is called a \emph{true zero} when  $K_{ij}=0$ and \emph{false zero} when $K_{ij} \neq 0$. If $\widehat{K}_{ij} \neq 0$, it is called a \emph{true nonzero} when  $K_{ij} \neq 0$ and $\s(\widehat{K}_{ij}) = \s(K_{ij})$, a \emph{false nonzero} when  $K_{ij} = 0$ and a \emph{wrong sign} when $K_{ij} \neq 0$ and $\s(\widehat{K}_{ij}) \neq \s(K_{ij})$.

Our priorities are reflected in the discrepancy measure matrix $M$ defined by the squared relative difference
\begin{equation}
   M_{ij} = M_{ij}(\widehat K,K)= \frac{(\widehat K_{ij} - K_{ij})^{2}} {(|\widehat K_{ij}| + |K_{ij}|)^{2} + \epsilon}, \quad i,j \in N.
\label{e:mij}
\end{equation}
The very small positive number $\epsilon$, much less than the desired accuracy, is included to make the definition valid also if both elements are zero. It is a matter of elementary algebra to show that $M_{ij}$ satisfies the requirements of a distance measure in $R$. Obviously,  $M_{ij} = 0$ for a correct estimate, while $M_{ij} \approx 1$ for a false nonzero or a false zero in $\widehat K_{ij}$ or if $\widehat K_{ij}$ comes with the wrong sign. In all other cases, $0 < M_{ij} < 1$, a small value indicating a good estimate. If the error is small, $M_{ij}$ is approximately equal to half the relative error in $\widehat J_{ij}$. The average discrepancy measure is
\begin{equation}
    \overline M(\widehat K,K) = \frac{1} {n^{2}}\sum_{i,j \in N}M_{ij}(\widehat K,K).
\label{e:m}
\end{equation}

\section*{Simulation results}
\label{simulationresults}

\subsection*{Estimating $J$ by modifying mRNA degradation rates}
\label{simmrnaproteingamma}

When the equilibrium concentrations are noise-free and $\widehat J$ is estimated by perturbing the mRNA degradation rates, it should in theory be equal to $J$ within computational accuracy, which depend on the user-defined relative perturbation $\omega$ of $\mu_{k}$, integration tolerance, etc. Numerical simulations on randomly generated gene regulatory networks (see below) show that choosing $\omega$ of the order $10^{-2}$ to $10^{-3}$ gives a discrepancy measure of the same order of magnitude. By reducing $\omega$ sufficiently, $\overline M$ can in theory be brought down to zero.

With real data, however, this is unattainable due to noise and experimental inaccuracy. Too small values of $\omega$ will lead to large uncertainties in the estimates of the derivatives. To obtain better estimates for the derivatives a fairly large number of observations with different perturbation levels would be needed. 
Many methods to estimate derivatives from noisy data can be found in the literature.  Considering this to be part of the experimental and data processing setup, we do not elaborate this point any further. Instead, we limit ourselves to assuming that a single perturbation level is used, and that this experiment is repeated a certain number of times to produce a distribution of observed values. With this approach one must seek an optimal tradeoff between reducing the error in the derivatives due to finite differences (using a small $\omega$) and minimising the noise to signal ratio. The differences in the equilibrium concentrations should not vary too much due to the noise, while $\omega$ should not be so large that nonlinear effects jeopardise the estimates of the derivatives.  Intuitively, the perturbation level $\omega$ should be considerably larger than the standard deviation of the distributions of equilibrium levels.

\subsection*{Estimating Jacobian for the segment polarity network by means of allele knockouts}
\label{simsegmpol}

With its crude estimate of the derivatives it is less obvious that the allele knockout approach will work with an acceptable accuracy. To test this, we applied this method to the single cell segment polarity network \citep{Tegner2003}. We computed the protein equilibrium values by integrating the rate equations until the stable state $x$ was reached with high accuracy. Employing total least squares, we used the equilibrium data to compute a matrix $\widehat J^{0}$ from Eq.~\eqref{e:jest2}, then computed $\widehat K^{0} = \widehat J^{0} + \Gamma$, and finally subjected $\widehat K^{0}$ to  two kinds of cutoff to arrive at our final estimate $\widehat K$ (see Methods for details). To simulate the effect of noisy data, we also repeated the computation of $\widehat K$ after uniform noise had been added to $x$.  More precisely, before estimating $\widehat K$ we added a noise term $L u_{i} x_{i}$ to each equilibrium concentration $x_{i}$, where $u_{i}$ was uniformly distributed in $[-1,1]$ and $L$ increased in steps of 0.05 from 0 to 0.25. For each noise level we ran $\ell = 50$ simulations. 

\begin{figure}[h!t]
\begin{center}
\includegraphics[width=2.5in]{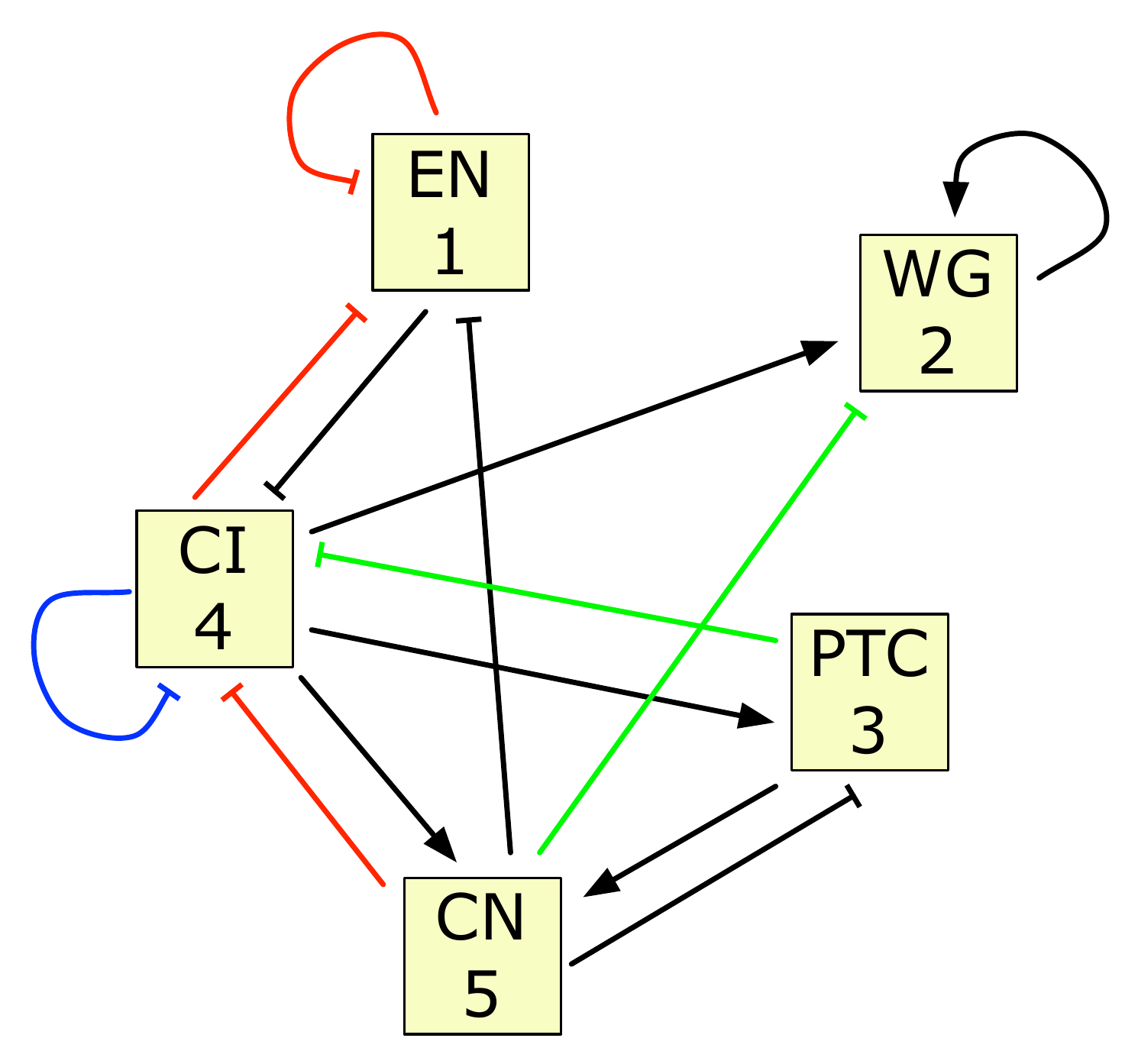}
\end{center}
\caption{The protein connectivity network for the single cell segment polarity network according to \citet{Tegner2003}. Arrow heads signify activation, crossbars inhibition. Colour codes refer to our estimated $\widehat K$ without noise. Black arcs are correctly predicted in $\widehat K$ with the right sign, blue arc are predicted with the wrong sign, green arcs signify false zeros (predicting no arc where there is one), and red arcs signify false nonzeros (predicting an arc where there is none). Thus, in the correct graph, black, blue and green arcs should be included, red arcs excluded. }
\label{segmpolnet}
\end{figure}

The true and the predicted network connections obtained when no noise has been added to the protein equilibrium data, are shown in Figure \ref{segmpolnet}. The cutoff value $c_{\mathrm J} = 0.003$  appeared to give the lowest number of false elements.  The elements are colour coded to show false nonzero elements (red), false zero elements  (green) and nonzero but false sign (blue). With the same colour coding the discrepancy measure is
\begin{equation}
   M = \left(\begin{array}{ccccc}
    {\color{red}1}      &   0   &      0  &  {\color{red}1}    & 0.032\\
         0   & 0.015    &     0 &   0.009  &  {\color{green}1} \\
         0     &    0    &     0  &  0.002  &  0.001\\
    0.206    &     0   & {\color{green}1}    & {\color{blue}1}   &  {\color{red}1} \\
         0    &     0  & 0.014   & 0.001   &      0
         \end{array}\right)
\label{e:m00}
\end{equation}
with average value $\overline M = 0.251$.  For noisy data with noise level up to 0.25 the results are similar, with roughly the same average discrepancy measure (see Appendix \ref{segmpol}). Apart from the false nonzeros, false zeros and false signs for which $M_{jk} = 1$,  the estimates are all of the right order of magnitude.

\subsection*{Estimating Jacobian for randomly generated gene regulatory networks by allele knockouts}
\label{simgrn}

The segment polarity network model has fixed network structure and parameter values. To complement this we performed large-scale simulations of gene regulatory network models with varying number of genes ($n=5,10,20$). For each network size we ran 100 Monte Carlo simulations, sampling network connectivities and wild-type parameter values. The model structure and simulation setup is explained in detail in the Methods section. For each of the 100 randomly generated systems we simulated $\ell =25$ sets of single knockout experiments, added noise ($L=0,0.05,0.1,0.15$) to the  steady state expression levels and computed $\widehat K$ by means of ordinary least squares, trimming the estimates by the cutoffs $c_{\mathrm J}=0.005$ and $c_{\mathrm S}=0.5$ (see Methods).

\begin{figure}[htb]
\centering
\scalebox{0.6}{\includegraphics{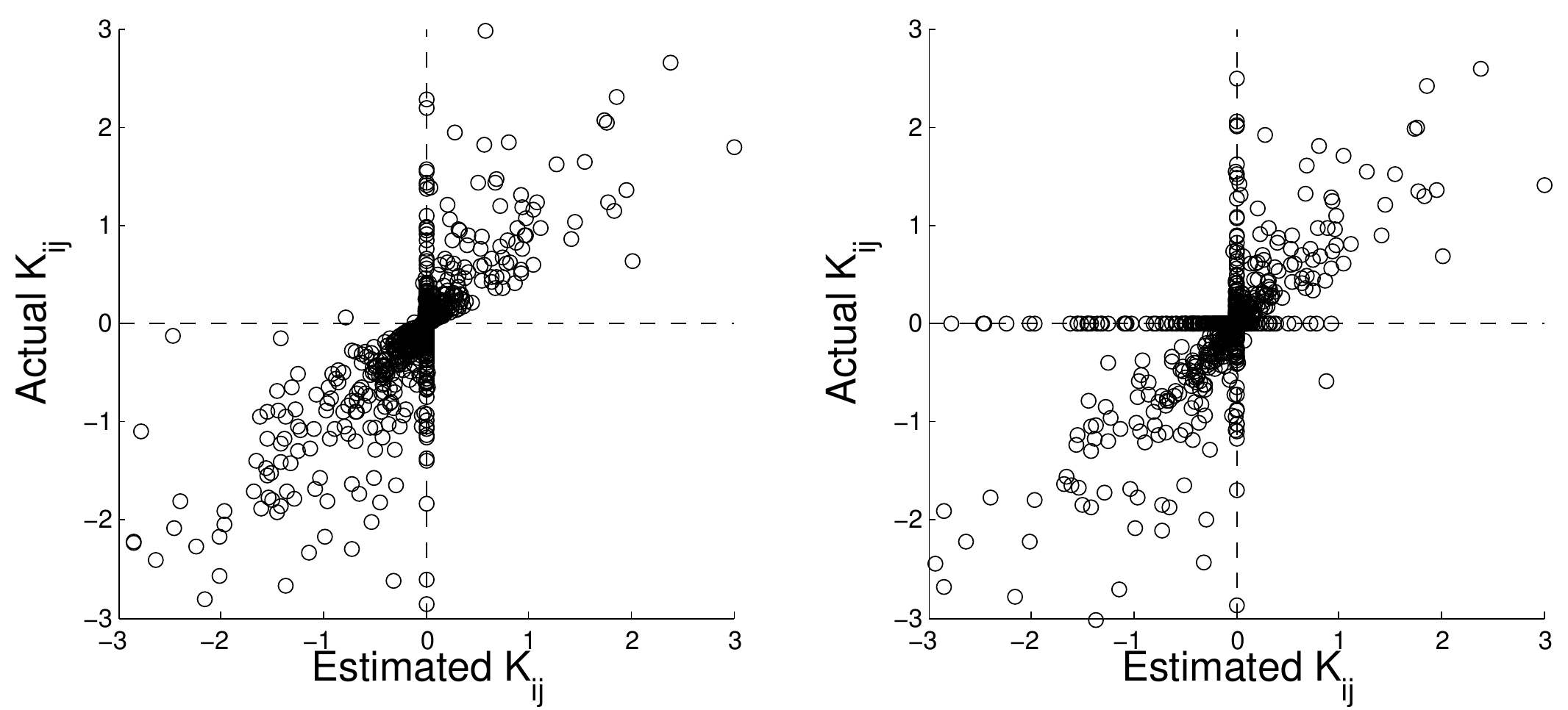}}
\caption{Scatterplots of $\widehat{K}_{ij}$ (x-axis) versus $K_{ij}$ (y-axis) for \emph{in silico} single-knockout experiments on 100 gene regulatory systems with $n=10$ genes. The left panel shows results without noise on steady state expression levels while the right panel shows results from $l=25$ repeated measurements with noise level $L=0.1$. Observations where $\max\{|\widehat{K}_{ij}|,|K_{ij}|\}>3$ are not shown (68 and 85 of 10000 cases for the left and right panels, respectively).}
\label{scatterplots}
\end{figure}

Figure \ref{scatterplots} shows scatterplots of estimated versus true values of elements in $K$ for $n = 10$. For noise-free steady-state expression levels (left panel) we observe mainly \emph{true nonzeros} in the 1st and 3rd quadrants and \emph{false zeros} on the $K_{ij}$-axis. When noise is added to the steady-state expression levels (right panel), \emph{false nonzeros} appear on the $\widehat K_{ij}$-axis. Estimates with \emph{wrong sign} would appear in the 2nd and 4th quadrants, but are rarely seen. Similar patterns are observed for systems with $n=5$ and $n=20$ (see Appendix \ref{randomly}).

\begin{figure}[h!tb]
\centering
\scalebox{0.35}{\includegraphics{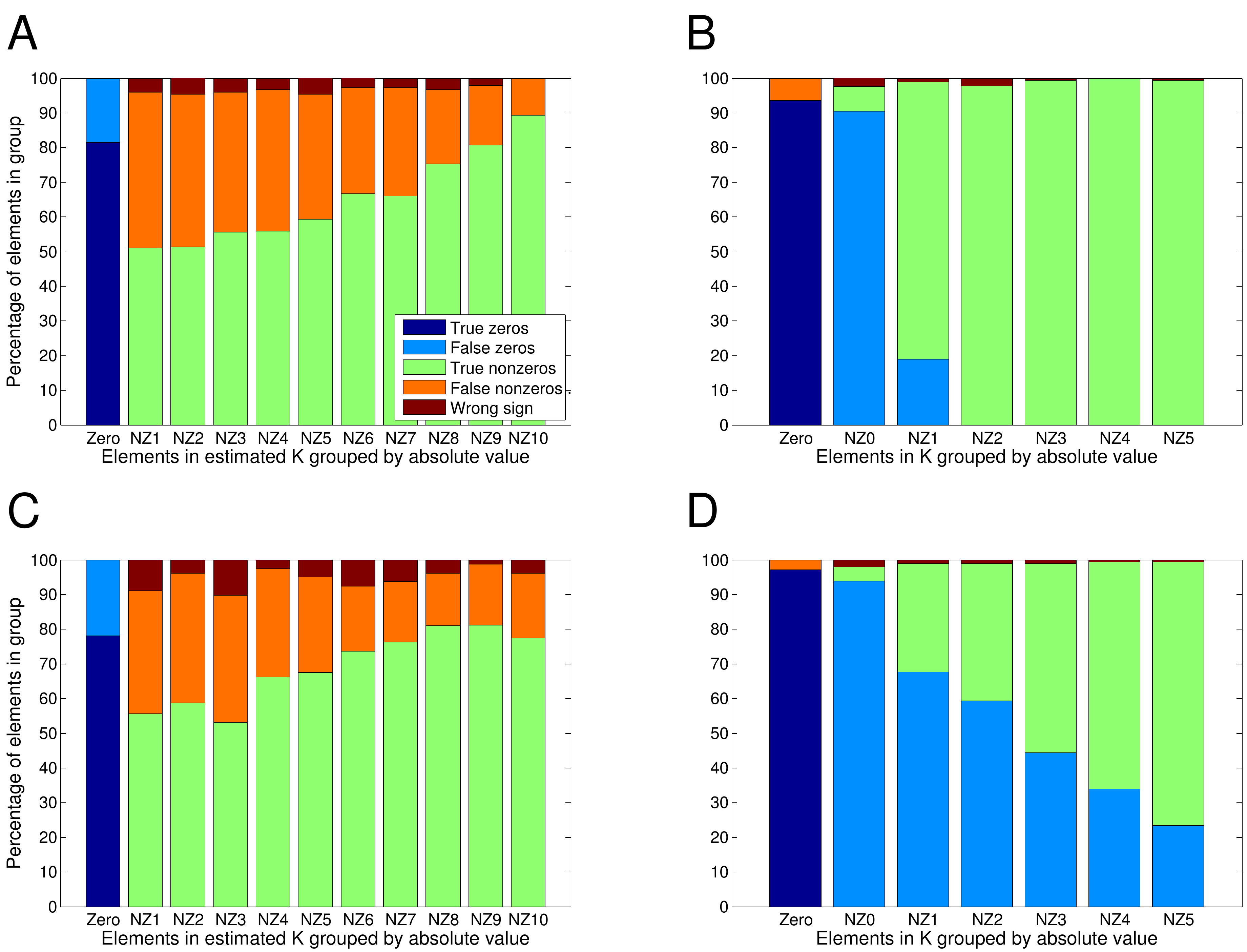}}
\caption{Summary of true and false discoveries of the signs of Jacobi elements for randomly generated gene regulatory networks with $n = 10$ genes. Each panel summarizes 10,000 $(\widehat{K}_{ij},K_{ij})$ pairs from \emph{in silico} single-knockout experiments on 100 simulated systems. \textbf{(A)} Results for simulations without noise on steady state expression levels. The $(\widehat{K}_{ij},K_{ij})$ pairs are sorted into subsets (x-axis) based on $|\widehat{K}_{ij}|$. The subset named Zero contains 8,495 pairs with $|\widehat{K}_{ij}|=0$, while the remaining pairs are sorted into 10 subsets NZ$p$, $p=1,2,\ldots,10$, with boundaries corresponding to the $(p-1)$th and $p$th 10-quantiles of the 1,505 $|\widehat{K}_{ij}|$ values.  \textbf{(B)} Results for simulations without noise on steady state expression levels. The $(\widehat{K}_{ij},K_{ij})$ pairs are sorted into subsets (x-axis) based on $|K_{ij}|$. The subset Zero contains 7,394 pairs with $|K_{ij}|=0$, while the remaining pairs are sorted into 6 subsets; NZ0 which contains elements 1706 with $|K_{ij}| \leq c_J$ and  NZ$p$, $p=1,2,\ldots,5$, with boundaries corresponding to the $(p-1)$th and $p$th 5-quantiles of the remaining 900 elements with $|K_{ij}| > c_J$. \textbf{(C)} Results for simulations with noise level $L=0.1$ on steady state expression levels. The $(\widehat{K}_{ij},K_{ij})$ pairs are sorted into subsets (x-axis) based on $|\widehat{K}_{ij}|$, as for \textbf{(A)}. The Zero subset contains 9,202 pairs. \textbf{(D)} Results for simulations with noise level $L=0.1$ on steady state expression levels. The sorting of pairs is the same as in \textbf{(B)}. 
}
 \label{barplots}
\end{figure}

Not visible in the origin of Figure \ref{scatterplots}  is a large number of \emph{true zeros}, and the figure does not convey much information about the relative numbers of true and false sign estimates. Figure \ref{barplots} gives an overview of this by subdividing the pairs of $(\widehat{K}_{ij},K_{ij})$ values into subsets based on their absolute values and displaying relative number of true and false positives for each subset. Using our methods on noise-free steady-state expression levels we find that slightly over 80\% of the cases where $\widehat{K}_{ij}=0$ are \emph{true zeros}, while the proportion of non-zero $\widehat{K}_{ij}$ that are \emph{true non-zeros} increase with $|\widehat{K}_{ij}|$ from ~50\% for the smallest estimates to ~90\% for the largest (Figure \ref{barplots}A). When dividing $(\widehat{K}_{ij},K_{ij})$ pairs into subsets based on $K_{ij}$ (Figure \ref{barplots}B), we find that more than 90\% of the zero elements in $K$ are correctly identified as zeros. Furthermore, for the nonzero elements in $K$, false zeros are only a major problem in the group NZ0 where $|K_{ij}| \leq c_J$. The most important effects of adding noise to the steady-state expression values  is (i) a considerable increase in the proportion of large elements in $K$ that are incorrectly identified as zeros (Figure \ref{barplots}D, NZ1 to NZ5) and (ii) a reduction in the number of false non-zeros (Figure \ref{barplots}C, D). The overall pattern is similar for systems with $n=5$ and $n=20$ (see Appendix \ref{randomly}).

The distribution of $\overline M$ is shown in Table 1. The values of $\overline M$ decrease with increasing number of nodes. The main reason is that the proportion of elements in $K$ that are zero increases, and as seen in Figure \ref{scatterplots} these zero elements are often correctly identified. For $n=5$, equilibrium values with a low noise-level ($L=0.1$) lead to slightly higher discrepancy measures, but for networks with 10 and 20 nodes there are no clear differences.

\begin{table}
\centering
\begin{tabular}{llllll}
\\ \\
\multicolumn{6}{p{4in}}{Table 1: Summary statistics of $\overline M$ for simulated allele knockouts in gene regulatory netw th varying number of genes ($n$) and noise level ($L$).}
\\ \hline \\
$n$ & $L$ & $min(\overline M)$ & $median(\overline M)$ & $mean(\overline M)$ & $max(\overline M)$    \\
\hline
5  & 0   & 0.177  & 0.384  & 0.393  & 0.671  \\
5  & 0.1 & 0.233  & 0.421  & 0.423  & 0.77   \\
10 & 0   & 0.165  & 0.236  & 0.242  & 0.394  \\
10 & 0.1 & 0.173  & 0.241  & 0.244  & 0.363  \\
20 & 0   & 0.09   & 0.133  & 0.137  & 0.229  \\
20 & 0.1 & 0.102  & 0.137 & 0.139 & 0.182  \\
\hline
\\ 
\label{tableM}
\end{tabular}
\end{table}

\section*{Conclusions}
\label{conclusions}

We have presented a method for estimating the Jacobian of an ODE model of a dynamic gene regulatory system based on the stable equilibrium values of the protein concentrations.  The method is developed from our previous analysis of propagation of genetic variation \citep{Plahte2013}. Together with known relative protein degradation rates, the observed shifts in the protein concentrations due to perturbations of the genes are sufficient to obtain an estimate $ \widehat J$ of the true Jacobian $J$. We have analysed two experimentally feasible ways of perturbing the genes: perturbing the relative mRNA degradation rates in haploid or diploid systems, and allele knockout in diploid systems. 

The reader should keep in mind that when one talks about a real system, the Jacobian always refers to a model of the system, not to the system itself. Our model framework is very general and includes explicit modelling of both mRNA and protein. We assume that the conversion rate from mRNA to protein is linear. However, as the Jacobian also is a linear representation valid around a stationary point of a usually nonlinear system, this linear conversion could be considered as a linear approximation of some nonlinear mRNA-protein response function.

Major advantages of our method are:
\begin{enumerate}
\item
The estimate $\widehat J$ can be obtained from Eq.~\eqref{e:jest1} or Eq.~\eqref{e:jest2}. Both equations are derived from Eq.~\eqref{e:jgq}. In this equation, $Q$ may be estimated from the shifts in equilibrium values obtained by \emph{any} kind of perturbation according to $q_{jk} \approx \delta_{j}^{[k]}/\delta_{k}^{[k]}$, c.f.~Eq.~\eqref{e:psiprimeest}. Knowing the relative degradation rates is not necessary. The effect of $C$ is to multiply each row in $H^{-1}$ by some factor proportional to the relative degradation rate $\gamma_{k}$ of the corresponding protein. Thus, varying $\gamma_{k}$ does not change the sign of the elements in $\widehat J$, only their magnitudes. Accordingly, all the connectivities except autoregulations can be estimated with their right sign even if the $\gamma_{k}$ are unknown.  We consider this  a major asset of the method.
\item If the protein relative degradation rates are known, $C$ may be estimated by modifying the mRNA relative degradation rates or by allele knockouts. In those cases $\widehat K$ can be computed, a non-negative value of  $\widehat K_{kk}$ pointing to an autoregulation in $X_{k}$.

\item
Already existing knowledge about the action of one node on itself or other nodes  can easily be incorporated.  For example, if $X_{j}$ is not autoregulated or negatively autoregulated, the value of $C_{jj}$  should be such that $\widehat J_{jj} < 0$. Only a sufficiently strong positive autoregulation may give $J_{jj} > 0$. Furthermore, if the sign of the action $X_{k} \to X_{j}$ is already known from experiments and the estimate of $(Q^{-1})_{j k}$ has the opposite sign, then necessarily $C_{jj} < 0$ if the sign of the estimate can be trusted. This fixes  the sign of all the other nonzero elements in row $j$ in $\widehat J$. Such knowledge may also be used to set admissible ranges for the cutoffs.
\end{enumerate}

Errors in $\widehat J$ could be due to observational noise or occur because derivatives have been approximated by ratios of finite differences. The effects of noisy data on the computation of $\widehat J$ have been analysed by \citet{Andrec2005}. Because $J$ is most likely sparse, in particular for large $n$, there will be a large number of false nonzero elements in $\widehat J$ before the cutoffs have been applied. If $J$ is known, as it were in our numeric simulations, the values of the cutoffs $c_{\mathrm J}$ and $c_{\mathrm S}$ can be chosen to obtain an optimal fit. Our sign fluctuation analysis in terms of $\overline S$ is of course quite simple-minded, and could be replaced  or supplemented by more elaborate statistical analyses. In an experimental situation the optimal fit must instead be searched by repeated switching between experiment, theoretical analysis, and new, testable hypotheses based on high and low cutoff values. 

A minimum value of $c_{\mathrm J}$ can always be set because too small values of $\widehat J_{jk}$ will be interpreted as indicating no effective regulatory action. However, the best cutoff value may be larger than this minimum. 
Large cutoffs gives fewer false nonzeros, but more false zeros (predicting a zero element in $\widehat J$ where the corresponding value in $J$ is nonzero), and vice versa. Setting large cutoffs will lead to just a few nonzero elements in $\widehat J$ which most likely are correct, but a large number of false zeros. Similarly, setting small cutoffs will lead to near-certain zeros in $ \widehat J$, but many false nonzeros. The optimal cutoff values are a trade-off between these two extremes.

Even if the true Jacobian is unknown, the estimate $\widehat J$ could be subjected to a few tests. All the eigenvalues of the optimal $ \widehat J$ should have negative real parts. This is not a strong criterion because the negative diagonal elements stemming from the degradation terms tend to make $\widehat J$ stable. A second consistency check is to use the sign conditions Eq.~\eqref{e:fdelta} relating the sign of the allele interaction values to the dominant loops. 

Estimating derivatives from noisy data is notoriously difficult and error-prone. However, if the genes may be perturbed by small amounts and with many different values (for instance by using RNA interference), more elaborate methods could be used to estimate the derivatives $q_{jk}$, even when data are noisy. The optimal method would depend on available experimental data, and is therefore outside the scope of this paper. Allele knockout, on the other hand, does not admit these options. An allele is either present or knocked out, leaving the researcher with just two data points from which the derivative may be approximated by finite differences. Accordingly, allele knockout is inherently a less precise method than degradation rate perturbation. For this reason, our simulations and discussion are mainly devoted to this suboptimal method in an attempt to determine its potential. Of course, combinations of the two methods could also be  envisaged. Thus, our statistical analyses of the equilibrium data should be considered more as examples than recipes on how the data should be analysed. Our object has not been to develop the optimal way of analysing specific data, but to illustrate different approaches and to show that our method actually works. 

Using steady state data, our method gives a sign estimate of the Jacobian, and if  the protein degradation rates are known, a value estimate. Some connections are predicted more reliably than others, and should point to further experiments that may resolve the inconsistencies and uncertainties in the more uncertain estimates. However, with real data it is not to be expected that one reconstruction method alone will give the complete and correct answer. Where one methods is inaccurate or fails, another may work, perhaps leading to conflicting conclusions that have to be resolved by further experimentation and analysis.

\section*{Networks and models}
\label{methods}

We tested both approaches for $J$ reconstruction---allele knockouts and degradation rate manipulation---on \emph{in silico} networks. In all the networks considered, protein and mRNA are modelled separately according to our general model framework. The stable protein concentration values were obtained by numeric integration of the rate equations until convergence to a steady state. In a few cases the solution approached a limit cycle. These cases were discarded.
We compared  $\widehat J$ estimated by the approaches described above to the ``true'' Jacobian $J$ estimated directly from the rate equations in the standard numeric way by estimating the partial derivatives of the rate functions in the reduced model Eq.~\eqref{e:reduced} (without the factor $2$ if the system is not diploid).

\subsection*{Segment polarity network}
\label{segment}
One of the systems we used to test the method is the segment polarity network model of \citet{vonDassow2000} as it was adapted to a single cell \citep{Tegner2003}. In this model $m_{i}$ and $P_{i}$ are the mRNA and protein concentrations of the genes \emph{engrailed (en)} ($i = 1$), \emph{wingless (Wnt)} ($i = 2$), \emph{Patched (Ptc)} ($i = 3$), \emph{cubitus interruptus (ci)} ($i = 4$), and repressor fragment of \emph{cubitus interruptus} ($i = 5$), respectively. Application of the quasi-stationarity hypothesis to the equations of motion given in  \citet[see Supplement]{Tegner2003}, leads to a realisation of Eqs.~\eqref{e:reduced} for all $i \neq 4$ and 
\begin{equation}
	\dot z_{4}  = \frac{\lambda_{4}}{\mu_{4}}r_{4}(z) - r_{5}(z) - \gamma_{4}z_{4}.
\label{e:p4}
\end{equation}
Here $z \in R^{5}$ is the vector of protein concentrations, $r_{i}$ are the mRNA dose-response functions, and $\lambda_{i}$, $\gamma_{i}$, and $\mu_{i}$ are constant parameters (see \citet{Tegner2003} for explicit formulae, parameter values and other details). Due to the presence of $r_{5}$ in the equation for $\dot z_{4}$, this system does not fit completely with our assumptions. A genotypic variation in $X_{5}$ will affect the dose-response function of $z_{4}$ directly, implying that the parameters describing the genotype of $X_{5}$ are not completely node-specific. It is interesting to see whether this fact will jeopardise our reconstruction or the system's Jacobian.

\subsection*{Random network models}
\label{random}
As a further test we ran a series of numerical simulations with and without noise on a range of systems defined by Eqs.~\eqref{e:reduced2} of varying dimension and feedback structure for randomly generated dose-response functions and parameter values. For each system size ($n=5,10,20$) we sampled 100 systems. For each system we first set up the overall connectivity as follow: for each node $X_i$ we sampled two regulator nodes $X_j,X_k$, $i,j,k \in N$, and a mode of regulation (activation or repression) for each regulator. This simplifies the rate equation  Eq.~\eqref{e:reduced2} to 
\begin{equation}
   \dot{z}_{i} =  2\frac{\rho_i}{\mu_i}R_{i}(z_j,z_k,a_{i}) - \gamma_{i}z_{i}. 
\label{e:simsystem}
\end{equation}

We used the dose-response function 
\begin{equation}
 	R_{i}(z_j,z_k)=\alpha_i+\beta_i B_i(S_{ij}(z_j),S_{ik}(z_k)),
	\label{e:simgrf}
\end{equation}
where $\alpha_i$ is the basal and $\alpha_i + \beta_i$ the maximal mRNA production rate, and $B_i$ is the algebraic equivalent of a Boolean AND or OR function. We set $S_{ij}(z_j) = H(z_j,\theta_{ij},p_{ij})$ if $X_j$ activates $X_i$ and $S_{ij}(z_j) = 1 - H(z_j,\theta_{ij},p_{ij})$ if $X_j$ represses $X_i$, where $H$ is the Hill function $H(z,\theta,p) = z^p/(z^p+\theta^p)$ with threshold $\theta$ and steepness $p$.
We sampled parameter values uniformly in the following ranges: 
$\beta_i, \mu_i \in (0,10)$,
$\alpha_i, \rho_i, \gamma_i, \theta_{ij}, \theta_{ik} \in (0,1)$ and $p_{ij},p_{ik} \in  (1,5)$.

\section*{Estimation methods}
\label{leastsquares}

\subsection*{Least squares}
The question of how to work out the best estimate $\widehat J$ from allele knockout or degradation rate perturbation data is not trivial. In all cases we get conditions of the kind $\widehat JH = G$, where $H$ and $G$ are square and $G$ is diagonal. If there is just a single set of measurements and $H$ is invertible, our estimated Jacobian is uniquely given by $\widehat J = GH^{-1}$. If data are noisy and we have a set of  observations leading to $H_{i}$ and $G_{i}$, $i    = 1,\ldots, \ell$, it is a matter of statistics to decide on an optimal estimation procedure. Obvious options are ordinary least squares (OLS) and total least squares (TLS) \citep{Markovsky2007}.  OLS assumes no measurement error in $H$, but works well as long as the measurement errors are small \citep{Montgomery2012}. In its simple form, TLS assumes equal variances in $H$ and $G$, which is at best only approximately fulfilled in our case. We used TLS for the segment polarity network and OLS for the randomly generated systems. (See Appendix \ref{ols} for details on OLS and TLS.) We do not claim that these are the optimal estimation procedures. Rather, they should be considered as examples used to show that the method actually works. More sophisticated estimates could certainly be found, but considering this as a problem belonging to the experimental and data processing setup, we do not elaborate this point any further.

\subsection*{Cutoffs}
Since real networks, in particular large ones, seem to have sparse Jacobians, $\widehat J$ will in general probably contain many elements with small absolute values. The question is then whether these are just the consequence of approximation errors in the estimation procedure, or correspond to weak couplings in the network. Our algorithm does not make any qualitative distinction between an element in the Jacobian with a small absolute value and a zero element. This is reasonable in light of the probabilistic nature of transcription suggested by \citet{Bintu2005}, who express the value of the dose-response function by the binding probabilities of transcription factors and polymerase. However, in common deterministic models an action of one gene on another either exists or does not exist. To relate our estimated $\widehat J$ to this kind of models we therefore have to apply some kind of cutoff to small elements in $\widehat J^{0}$, the estimate without cutoffs computed by one of the above estimation procedures..

In our simulations we applied two cutoffs to $\widehat K^{0} = \widehat J^{0} + \Gamma$. First we set all elements in $\widehat  K^{0}$ with an absolute value less than a cutoff $c_{\mathrm J}$ to zero. Secondly, for the cases of noisy data, we used the estimates $\widehat K_{l}^{0} = \widehat J_{l}^{0} + \Gamma$ obtained by means of Eq.~\eqref{e:jest1} or Eq.~\eqref{e:jest2} for all $\ell$ data set to define  the average sign matrix $\overline S = (1/\ell)\sum_{l=1}^{\ell}\mathrm{sign}(\widehat K_{l}^{0})$. The elements $(\widehat K_{l}^{0})_{ij}$ that are consistently equal to zero or have strongly varying sign for varying $l$, will come out as zero in $\overline S$ or with small absolute values. If $|\overline S_{ij}|$ is smaller than a chosen sign variation cutoff $c_{\mathrm S}$, the corresponding element in $K$ is most likely zero. In these cases we set $\widehat K_{ij} = 0$. In simulation studies when the true $J$ is known,  an optimal $c_{\mathrm S}$ can be found by comparing $\widehat K$ with $K$ for different $c_{\mathrm S}$. In real cases, a combination of intuition and repeated experiments may help in choosing the optimal value.

\section*{Competing interests}
The authors declare that they have no competing interests.

\section*{Authors' contributions}
EP conceived the study and carried out the mathematical analysis. ABG designed and ran the numerical simulations and performed the statistical analysis. Both authors took part in evaluating the results and drafting the manuscript, and approved the final version. 

\section*{Acknowledgements}
We thank Stig W.~Omholt for encouragement and comments. This work has been supported in part by The Research Council of Norway, project number 178901/V30, ``Bridging the gap:~disclosure, understanding and exploitation of the genotype-phenotype map''.

\section*{Appendices}

The appendices contain details of the singular perturbation analysis of the mRNA-protein system and of all derivations and details of the simulation procedure that are not 
included in the main document, and additional simulation results that are too lengthy to be included in the main document.

\numberwithin{equation}{section}

\appendix

\section{mRNA and protein networks}
\label{mrnaprotein}

The segment polarity network model of \citet{vonDassow2000} analysed in the paper is an example of a model framework in which the concentrations of mRNA and protein for each gene in the network are modelled independently. In this section we show how such models can be reduced by singular perturbation theory. For a network of $n$ genes the model framework is Eq.~\eqref{e:fullsystem}, repeated here for convenience:
\begin{equation*}
    \begin{split}
         \dot{P}_{i} & = \rho_{i} m_{i} - \sigma_{i} P_{i},\\
         \dot{m}_{i} & =  R_{i}(P) - \mu_{i} m_{i} .
            \end{split}
\end{equation*}
Here $P_{i}$ and $m_{i}$ are the concentrations of protein and mRNA of gene number $i$, respectively, $R_{i}$ is the production rate (dose-response function) of mRNA, dependent on the concentration of the input proteins,  $\rho_{i}$ is the mRNA-protein conversion rate, and $\sigma_{i}$ and $\mu_{i}$ are positive relative degradation rates. The gene products might act directly as transcription factors, or the function $R_{i}(P)$ might implicitly contain chains of reactions from the gene products to the real transcription factors so that $R_{i}$ is the combined effect of these chains and the transcription. This framework has been used by a number of authors, see e.g.~\citet{Polynikis2009} for a review.

As  mRNA molecules are in general less stable than the corresponding protein molecules, we can safely assume that for all $i$, $\sigma_{i} \ll \mu_{i}$. We define $\epsilon = \max\{\sigma_{i}/\mu_{i}\}$. Then $\epsilon \ll 1$, and by a suitable renumbering of the genes we can alway achieve $\epsilon = \sigma_{1}/\mu_{1}$.

Using  standard terminology in singular perturbation theory we call Eq.~\eqref{e:fullsystem} \emph{the full model}. To introduce $\epsilon$ in the equations we transform them to non-dimensional form by scaling the variables and the time $t$ according to 
\begin{equation}
\begin{split}
     	P_{i} & =  \frac{\rho_{i}}{\sigma_{i}\mu_{i}}x_{i},\\
	m_{i} & = \frac{1}{\mu_{i}}y_{i}.\\
	t & = \frac{1}{\sigma_{1}}T.
	\end{split}
	\label{e:scaling}
\end{equation}
For convenience we continue to use a dot to denote time derivatives, but now with respect to the scaled time $T$. This leads to the dimensionless equations
\begin{equation}
    \begin{split}
        \dot{x}_i & = \gamma_i y_{i} - \gamma_i x_i,\\
       \epsilon \dot{y}_i & = \eta_{i}R_{i}(x) - \eta_i y_i,
    \end{split}
\label{e:scaled}
\end{equation}
where $\eta_{i} = \mu_{i}/\mu_{1}$, $\gamma_{i} = \sigma_{i}/\sigma_{1}$.

When $\epsilon$ is small and with a number of reasonable assumptions, we can make the quasi-stationarity hypothesis 
\begin{equation}
   y_{i} = R_{i}(x).
\label{e:qshyp}
\end{equation}
This leads to \emph{the reduced model} 
\begin{equation}
\dot{x}_{i} = r_{i}(x) - \gamma_{i} x_{i},
\label{e:reduced3}
\end{equation}
where $r_{i}(x) = \gamma_{i}R_{i}(x)$. Obviously, the full and the reduced model have the same stationary states. The above derivation can be justified in a rigorous way by means of singular perturbation theory. In terms of the fast time $\tau = T/\epsilon$, Eqs.~\eqref{e:scaled} are
\begin{equation}
    \begin{split}
        x'_i & = \epsilon (\gamma_i y_{i} - \gamma_i x_i),\\
       y'_i & = \eta_{i}R_{i}(x) - \eta_i y_i,
    \end{split}
\label{e:fast}
\end{equation}
where the prime denotes differentiation with respect to $\tau$. The crucial necessary assumption for the singular perturbation to be valid is that the stationary point of Eq.~\eqref{e:fast} is asymptotically stable for  fixed values of $x_{i}$. In the present case this is ensured by assumption. Singular perturbation theory also ensures that when $\epsilon \to 0$, the solution of the reduced system approaches the solution of the full system for all $T$ except in a narrow, initial time interval.

\section{Estimate of $J$ by  least squares}
\label{ols}

In this section we consider briefly how to analyse the gene perturbation data using ordinary least squares (OLS) and total least squares (TLS). According to the main paper, both perturbation methods lead to 
\begin{equation}
   \widehat J_{l}H_{l} = a\Gamma B_{l} = A_{l},
\label{e:common}
\end{equation}
where the subscript $l = 1,\ldots,\ell$ indicates the data obtained from experiment number $l$, and $\Gamma$ is the diagonal matrix $\Gamma = \mathrm{diag}(\gamma_{k})$. For each $l$, $B_l = \mathrm{diag}(x_{k})$ in the case of degradation rate perturbation and $B_l = \mathrm{diag}(x_{k}^{[k]})$ in the case of allele knockout, $H$ is the matrix with elements $H_{jk} = x_{j} - x_{j}^{[k]}$, $a = - \omega$ in the case of degradation rate perturbation  and $a = -1$ in the case of allele knockout. Finally, $\widehat J_{l}$ is the Jacobian estimated from dataset number $l$, and $A_{l} = a\Gamma B_{l}$.

The problem is to derive the best fit to $J$ from the total dataset. Eq.~\eqref{e:common} can be rewritten as
\begin{equation}
   \Gamma ( \widehat J_{l}^{-1})^{\mathsf T} = \frac{1} {a}\left( H_{l}B_{l}^{-1}\right)^{\mathsf T}.
\label{e:rewritten}
\end{equation}
In this form the equation is amenable to an ordinary least squares solution because all noise is confined to the right-hand side. If $\Gamma$ also has a nonzero variance, this is no longer so, and a more careful analysis is necessary. Assuming  that $\Gamma$ is known with negligible inaccuracy, we form the system $Y = G( \widehat J^{-1})^{\mathsf T}$ by stacking all the right-hand sides of Eq.~\eqref{e:rewritten} into the $\ell n \times n$ matrix $Y$, and stacking $\ell$ $\Gamma$-matrices into the $\ell n \times n$ matrix $G$. Computing the least squares solution is straightforward and leads to
\begin{equation}
   \widehat J^{-1} = \frac{1} {\ell a}\sum_{l=1}^{\ell}\left( B_{l}^{-1}H_{l}^{\mathsf T}\right)\Gamma
^{-1}.
\label{e:invjhatsolution}
\end{equation}
Combining this with Eq.~\eqref{e:common} we readily arrive at
\begin{equation}
   \widehat J^{-1} = \frac{1} {\ell}\sum_{l = 1}^{\ell}\widehat J_{l}^{-1},
\label{e:finaljhat}
\end{equation}
from which we find our final estimate by inversion. 

The advantage of estimating $\widehat J^{-1}$ rather than $\widehat J$ is that the latter alternative would require the inverses of each $H_{l}$, while the above method essentially inverts the average of  all the $H_{l}$, which has a lower variance.

To compute the TLS solution we define $ H = [H_{1},H_{2},\ldots, H_{\ell}]$ and $A = [A_{1},A_{2},\ldots, A_{\ell}]$. Then Eq.~\eqref{e:common} can be written $JH = A + E$ or 
\begin{equation}
   H^{\mathsf T}J^{\mathsf T} = A^{\mathsf T} + E^{\mathsf T},
\label{e:hjq}
\end{equation}
where $E$ is the error matrix. Because there are uncertainties in both $H$ and $A$, TLS gives the optimal solution if all elements of $H$ and $A$ are normally distributed, uncorrelated and with equal variance. Assuming this is approximately correct, we find the TLS solution as follows \citep[Theorem 2]{Markovsky2007}. Let the dimensions of $H$ and $A$ be $n \times m$. We define $C$ by $C = [H^{\mathsf T},A^{\mathsf T}]$. Then the dimensions of $C$ are $m \times 2n$. We may assume that $m \geq 2n$. Let a SVD decomposition of $C$ be $C = USV^{\mathsf T}$, and let the singular values of $C$ be $\sigma_{1} \geq \ldots \geq \sigma_{2n}$. We partition $V$ as
\begin{equation}
   V = \left[\begin{array}{cc}V_{11} & V_{12} \\V_{21} & V_{22}\end{array}\right],
\label{e:v}
\end{equation}
each partition having dimensions $n \times n$. Then there exists a TLS solution if and only if $V_{22}$ is non-singular. In addition, the solution is unique if $\sigma_{n} \neq \sigma_{n+1}$. If both conditions are fulfilled, the TLS solution of Eq.~\eqref{e:hjq} is $\widehat J^{\ \mathsf T} = - V_{12}{V_{22}}^{-1}$, or
\begin{equation} 
  \widehat J = ({V_{22}}^{-1})^{\mathsf T}V_{12}^{\mathsf T}.
\label{e:widehatj}
\end{equation}

\section{Additional simulation results for the segment polarity network}
\label{segmpol}

The true and the predicted network connections obtained when no noise has been added to the protein equilibrium data, are shown in the main document. The true $K$ is
\begin{equation}
  K = \left(\begin{array}{ccccc}
       0     &    0    &     0     &    0  & -0.0562\\
         0   & 0.0412    &     0  &  0.0090  & -0.0036\\
         0     &   0    &     0   & 0.0168 &  -0.0066\\
   -0.0523    &     0  & -0.9650 &  -0.2565    &     0\\
         0     &    0   & 0.9650  &  0.2565     &    0
   \end{array}\right).
\label{e:segpolk}
\end{equation}
For $L = 0$ (no noise) we found
\begin{equation}
   \widehat K  = \left(\begin{array}{ccccc}
            {\color{red}-0.0160}      &    0    &     0  & {\color{red}-0.0153}   & -0.0806\\
         0  &  0.0527     &    0  &  0.0075    &     {\color{green}0} \\
         0    &     0     &    0  &  0.0184  & -0.0062\\
   -0.0196  &       0    &     {\color{green}0}  &  {\color{blue}0.0699}  &  {\color{red}-0.0327} \\
         0    &     0  &  1.2279   & 0.2381     &    0
 \end{array}\right),
\label{e:segmpoljhat00}
\end{equation}
using the cutoff value $c_{\mathrm J} = 0.003$ which appears to be optimal.  The elements are colour coded to show false nonzero elements (red), false zero elements  (green) and nonzero but false sign (blue). 

\begin{figure}[h!]
\centering
\scalebox{0.4}{\includegraphics{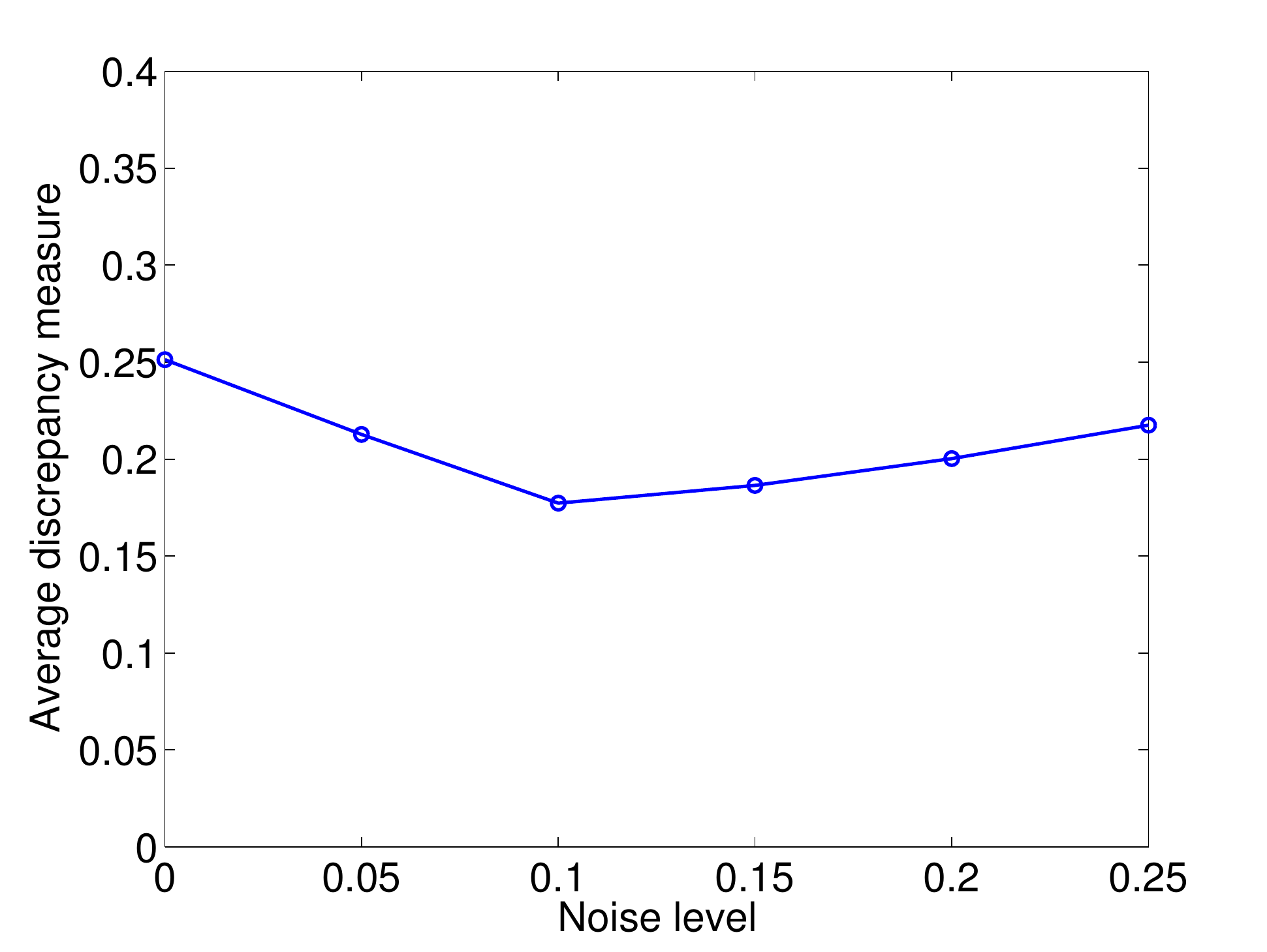}}
\caption{Average discrepancy measure $\overline M$ of the segment polarity network for noise level $L$ ranging from $L = 0$ to $L = 0.25$.}
\label{dmresults}
\end{figure}

With the same colour coding the discrepancy measure is
\begin{equation}
   M = \left(\begin{array}{ccccc}
    {\color{red}1}      &   0   &      0  &  {\color{red}1}    & 0.032\\
         0   & 0.015    &     0 &   0.009  &  {\color{green}1} \\
         0     &    0    &     0  &  0.002  &  0.001\\
    0.206    &     0   & {\color{green}1}    & {\color{blue}1}   &  {\color{red}1} \\
         0    &     0  & 0.014   & 0.001   &      0
         \end{array}\right)
\label{e:m00b}
\end{equation}
with average value $\overline M = 0.251$. Elements with false sign or false zeros/nonzeros in $\widehat K$ are equal to 1. For noisy data with noise level up to 0.25 the results are similar, with roughly the same average discrepancy measure, indeed somewhat smaller than without noise (Figure \ref{dmresults}). Apart from the false nonzeros, false zeros and false signs for which $M_{jk} = 1$,  the estimates are all of the right order of magnitude.

\begin{figure}[h!]
\centering
\scalebox{0.5}{\includegraphics{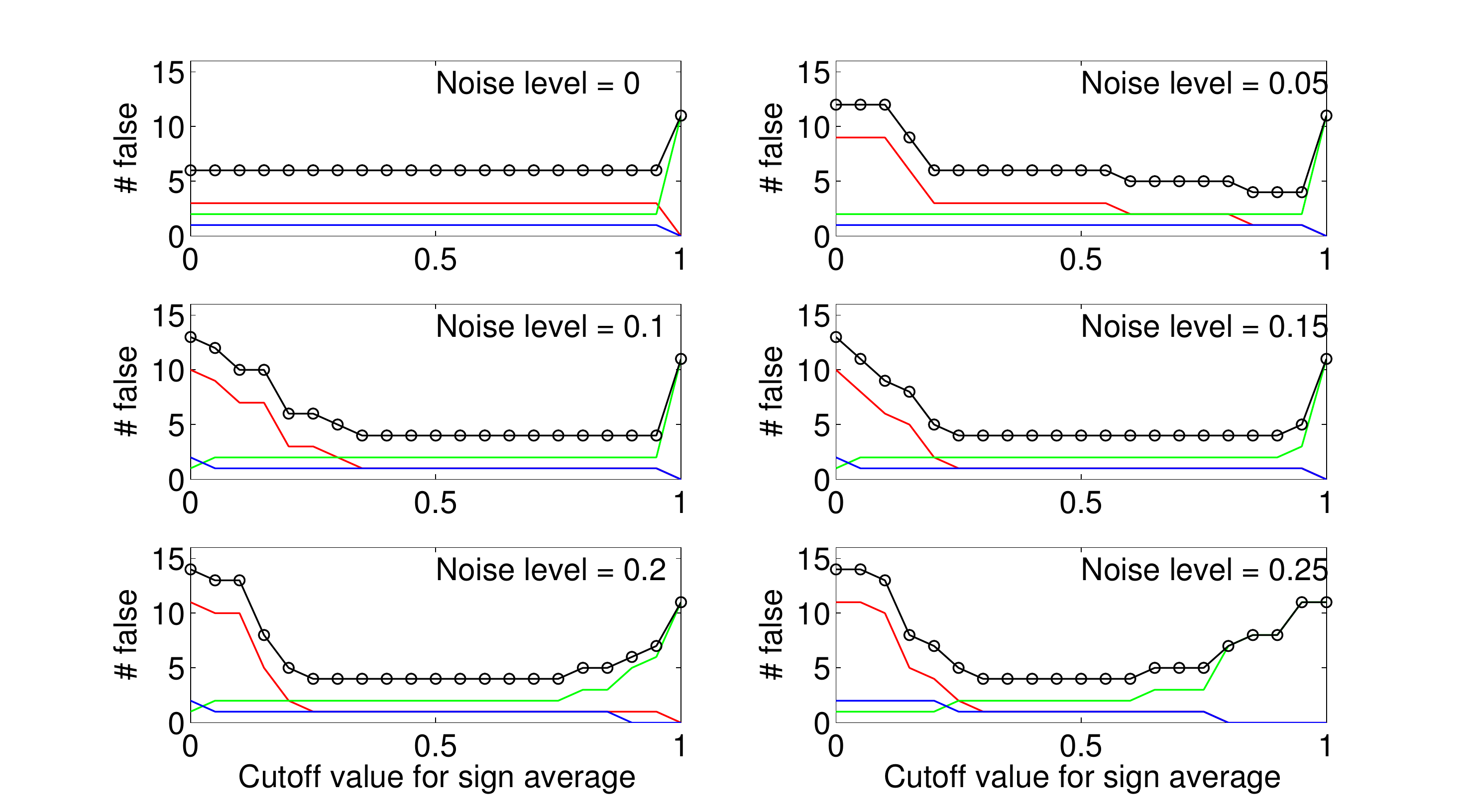}}
\caption{The number of false nonzero elements (red), false zero elements (green), false signs (blue) and the sum of all three (black with circles) in the estimated Jacobian for the segment polarity network for varying sign variation cutoff $c_{\mathrm S}$, and noise level $L$ ranging in steps of 0.05 from $L = 0$ (no noise) to $L = 0.25$.}
\label{segmpolnoise}
\end{figure}

\begin{figure}[h!]
\centering
\scalebox{0.5}{\includegraphics{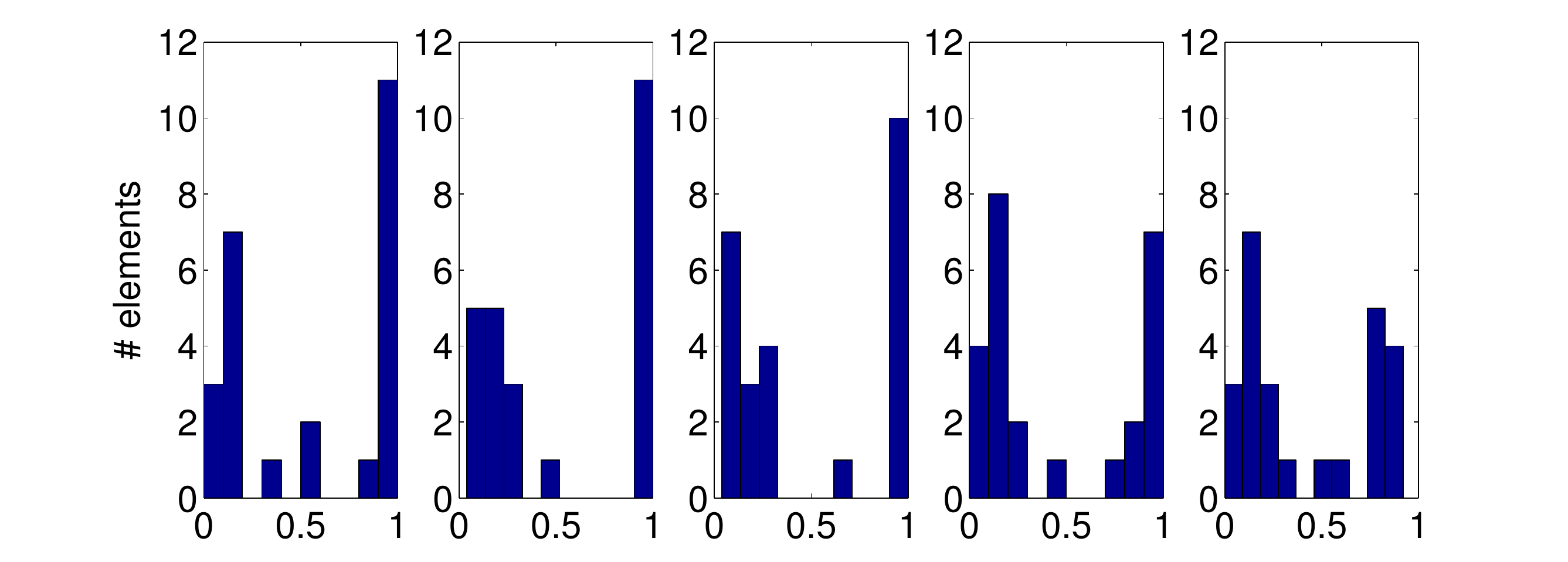}}
\caption{The distributions of the absolute values of the elements in $\overline S$ for noise levels $L = 0.05$ (far left) in steps of 0.05 to $L = 0.25$ (far right). }
\label{histograms}
\end{figure}

For $L = 0.10$ the average sign matrix $\overline S$ is
\begin{equation}
   \overline S  = \left(\begin{array}{ccccc}   
 {\color{red}-0.32}    & {\color{red}-0.20}  &   {\color{red}0.08}   & {\color{red}-0.28}    & -1  \\
    {\color{red}0.44}     & 1    & {\color{red}-0.08}    &  1    & -0.92  \\
    {\color{red}0.20}   & {\color{red}-0.24}    & {\color{red}0.08}   &   1   &  -1  \\
   -1    & {\color{red}-0.20}   &  0.04     & 1     & {\color{red}-1}   \\
    {\color{red}0.04}    & -{\color{red}0.20}  &   1   &  1   &  {\color{red}0.20} 
        \end{array}\right).
\label{e:s015}
\end{equation}
The elements that are zero in $K$ are shown  in red. With two exceptions, these are the elements whose values in  $\overline S$ are close to zero. With the same exceptions, the remaining elements in $\overline S$ have values close to $\pm 1$.  The same is true for the other noise levels investigated. For all noise levels, the magnitude of the elements in $\overline S$ are quite clearly separated in two classes, the smaller elements corresponding roughly to the fourteen zero elements in $K$ (Figure \ref{histograms}).

\section{Additional simulation results for the randomly generated systems}
\label{randomly}
In the main file we presented the results of the simulations on randomly generated systems with $n = 10$ genes. Here we present the corresponding results for $n = 5$ and $n = 20$. For interpretation of the diagrams see the legend to Figure \ref{barplots}.

\begin{figure}[h!]
\centering
\scalebox{0.6}{\includegraphics{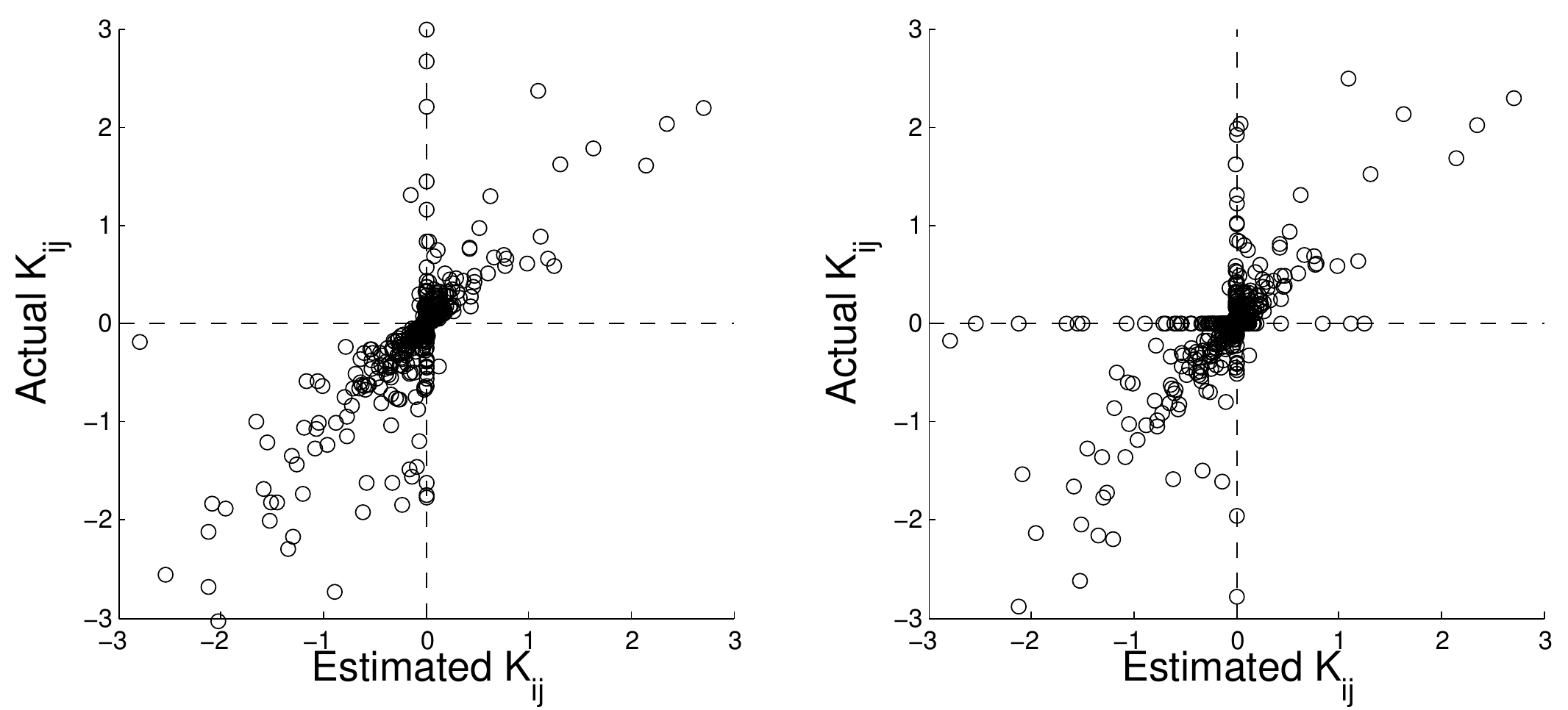}}
\caption{Scatterplots of $K_{ij}$ (y-axis) versus $\widehat K_{ij}$ (x-axis) for \emph{in silico} single-knockout experiments on 100 randomly generated gene regulatory systems with $n=5$ genes. The left panel shows results without noise on steady state expression levels. The right panel shows results from $\ell=25$ repeated measurements with noise level $L=0.1$. Observations where $\max(|\widehat{K}_{ij}|,|K_{ij}|)>3$ are not shown (24 and 26 of 2,500 ($\widehat{K}_{ij},K_{ij}$) pairs for the left and right panel, respectively).}
\label{s1}
\end{figure}

\begin{figure}[h!]
\centering
\scalebox{0.6}{\includegraphics{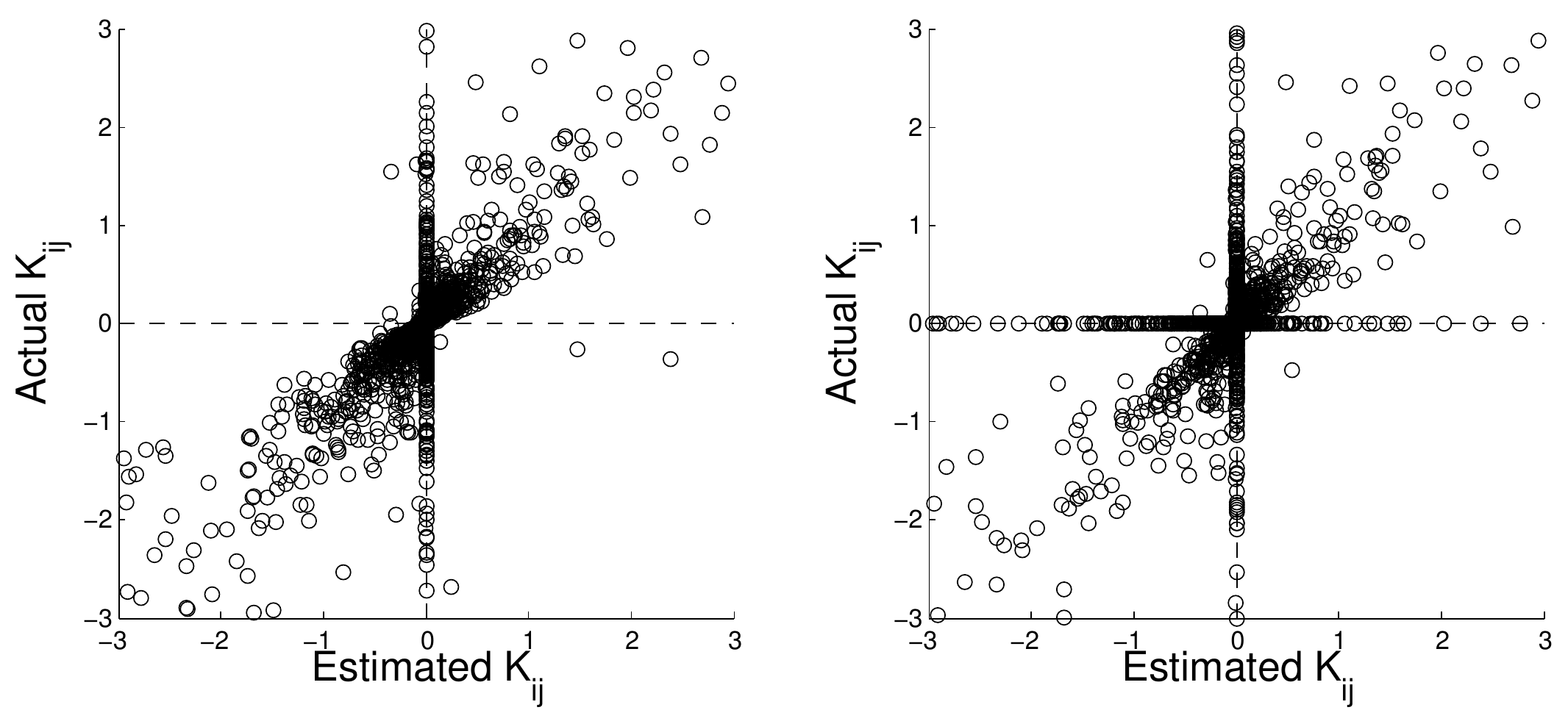}}
\caption{Scatterplots of $K_{ij}$ (y-axis) versus $\widehat K_{ij}$ (x-axis) for \emph{in silico} single-knockout experiments on 100 randomly generated gene regulatory systems with $n=20$ genes. The left panel shows results without noise on steady state expression levels. The right panel shows results from $\ell=25$ repeated measurements with noise level $L=0.1$. Observations where $\max(|\widehat{K}_{ij}|,|K_{ij}|)>3$ are not shown (157 and 174 of 40,000 ($\widehat{K}_{ij},K_{ij}$) pairs for the left and right panel, respectively).}
\label{s2}
\end{figure}

\begin{figure}[h!]
\centering
\scalebox{0.4}{\includegraphics{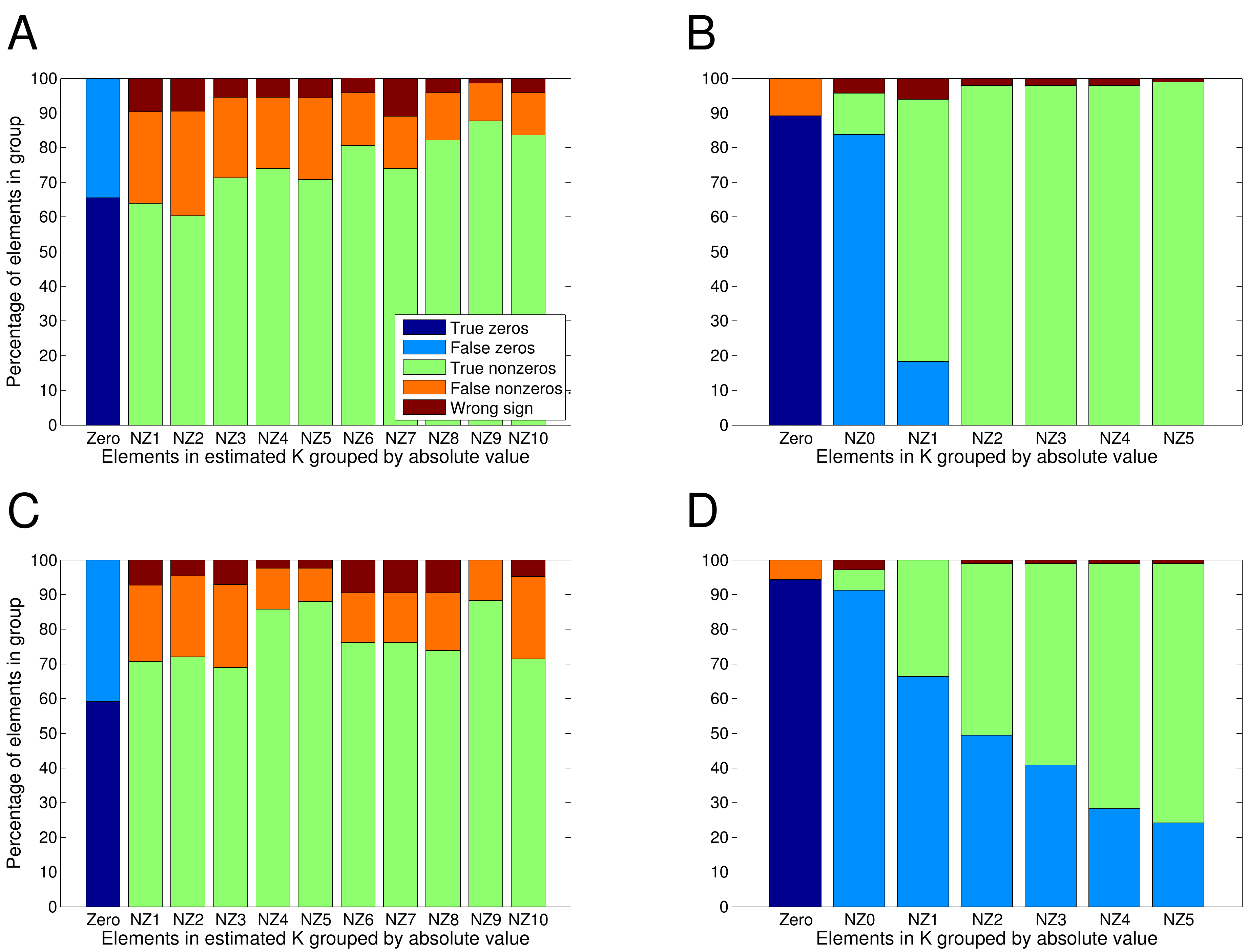}}
\caption{Summary of true and false discoveries of the signs of Jacobi elements  for randomly generated gene regulatory networks with $n = 5$. Each panel summarizes 2,500 $(\widehat{K}_{ij},K_{ij})$ pairs from \emph{in silico} single-knockout experiments on 100 simulated gene regulatory systems. \textbf{(A)} Results for simulations without noise on steady state expression levels. The $(\widehat{K}_{ij},K_{ij})$ pairs are sorted into subsets (x-axis) based on $|\widehat{K}_{ij}|$. The subset named Zero contains 1,771 pairs with $|\widehat{K}_{ij}|=0$, while the remaining pairs are sorted  into 10 subsets NZ$p$, $p=1,2,\ldots,10$, with boundaries corresponding to the $(p-1)$th and $p$th 10-quantiles of the 729 $|\widehat{K}_{ij}|$ values. \textbf{(B)} Results for simulations without noise on steady state expression levels. The $(\widehat{K}_{ij},K_{ij})$ pairs are sorted into subsets (x-axis) based on $|K_{ij}|$. The subset named Zero contains 2,077 pairs with $|K_{ij}|=0$, while the remaining pairs are sorted into 10 subsets NZ$p$, $p=1,2,\ldots,10$, with boundaries corresponding to the $(p-1)$th and $p$th 10-quantiles of the 423 $|K_{ij}|$ values. \textbf{(C)} Results for simulations with noise level $L=0.1$ on steady state expression levels. The $(\widehat{K}_{ij},K_{ij})$ pairs are sorted into subsets (x-axis) based on $|\widehat{K}_{ij}|$, as for \textbf{(A)}. The Zero group contains 2,006 pairs. \textbf{(D)} Results for simulations with noise level $L=0.1$ on steady state expression levels. The sorting of pairs is the same as in \textbf{(B)}.
}
\label{s3}
\end{figure}

\begin{figure}[h!]
\centering
\scalebox{0.4}{\includegraphics{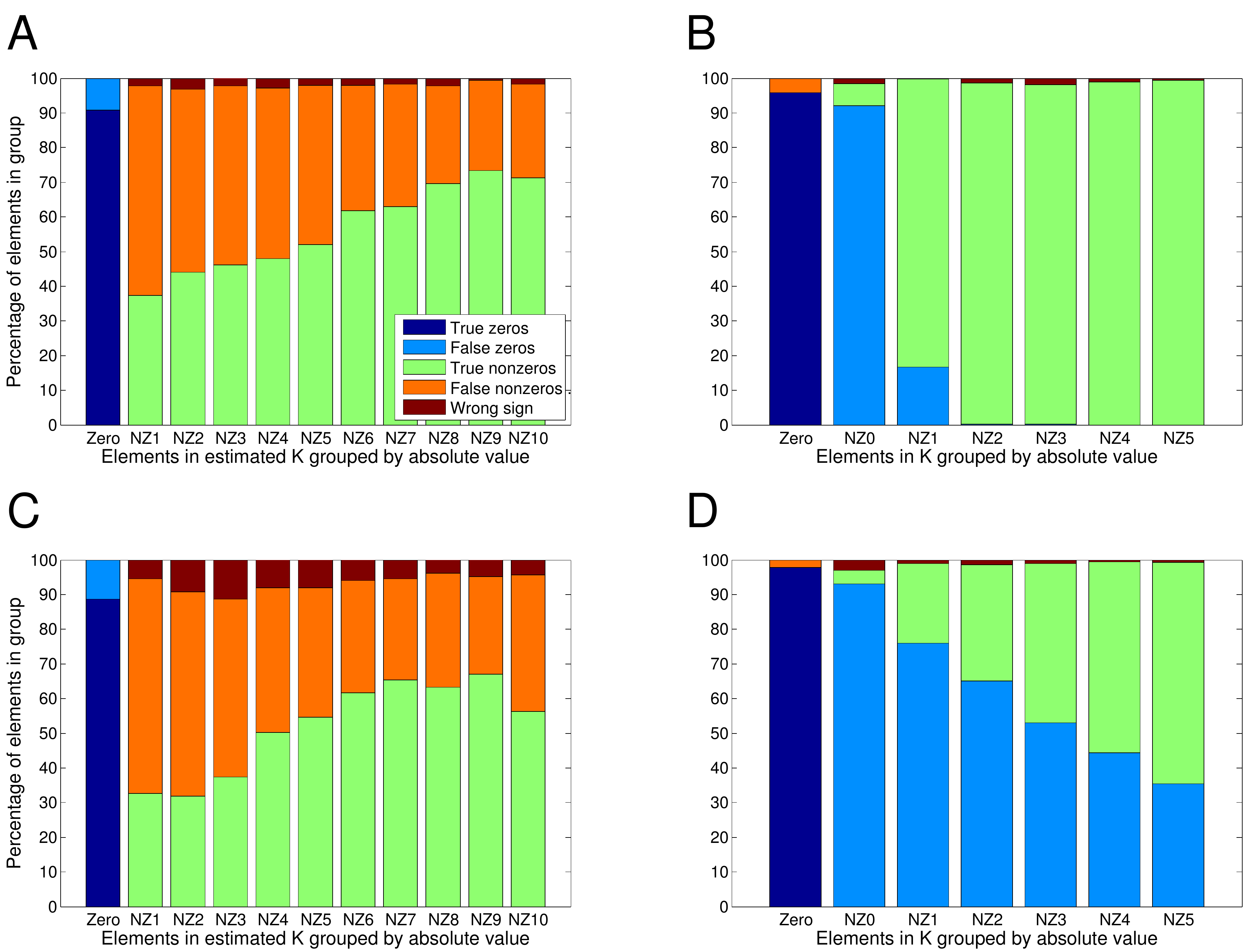}}
\caption{Summary of true and false discoveries of the signs of Jacobi elements for randomly generated gene regulatory networks with $n = 20$ genes. Each panel summarizes 40,000 $(\widehat{K}_{ij},K_{ij})$ pairs from \emph{in silico} single-knockout experiments on 100 simulated systems. \textbf{(A)} Results for simulations without noise on steady state expression levels. The $(\widehat{K}_{ij},K_{ij})$ pairs are sorted into subsets (x-axis) based on $|\widehat{K}_{ij}|$. The subset named Zero contains 36,462 pairs with $|\widehat{K}_{ij}|=0$, while the remaining pairs are sorted  into 10 subsets NZ$p$, $p=1,2,\ldots,10$, with boundaries corresponding to the $(p-1)$th and $p$th 10-quantiles of the 3,538 $|\widehat{K}_{ij}|$ values. \textbf{(B)} Results for simulations without noise on steady state expression levels. The $(\widehat{K}_{ij},K_{ij})$ pairs are sorted into subsets (x-axis) based on $|K_{ij}|$. The subset Zero contains 38,149 pairs with $|K_{ij}|=0$, while the remaining pairs are sorted into 10 subsets NZ$p$, $p=1,2,\ldots,10$, with boundaries corresponding to the $(p-1)$th and $p$th 10-quantiles of the 1,851 $|K_{ij}|$ values. \textbf{(C)} Results for simulations with noise on steady state expression levels. The $(\widehat{K}_{ij},K_{ij})$ pairs are sorted into subsets (x-axis) based on $|\widehat{K}_{ij}|$, as for \textbf{(A)}. The Zero subset  contains 38,139 pairs. \textbf{(D)} Results for simulations with noise level $L=0.1$ on steady state expression levels. The sorting of pairs is the same as in \textbf{(B)}.
}
\label{s4}
\end{figure}

When noise is added, the numbers of false elements also depend on the sign cutoff $c_{\mathrm S}$. However, for most noise levels, the value of $c_{\mathrm S}$ is not very critical.  Figure \ref{segmpolnoise} show the number of false nonzero elements, false zero elements, false signs and the sum of all three for varying $c_{\mathrm S}$ and a range of noise levels. For most noise levels the total number of false elements is four, corresponding to 84\% correctly predicted elements, for a wide range of cutoff values.

\newpage
\bibliographystyle{natbib}
 \bibliography{ReconstructGRN_refs}

\end{document}